\documentclass[twocolumn,prd,aps,superscriptaddress,preprintnumbers,tightenlines,showpacs,nofootinbib,amsfonts,amsmath]{revtex4-1}
\pdfoutput=1

\usepackage{color}
\usepackage{calc}
\usepackage{amsmath,amssymb,graphicx}
\usepackage{tensor}
\usepackage{bm}
\usepackage{times}
\usepackage[varg]{txfonts}
\usepackage[colorlinks, pdfborder={0 0 0}]{hyperref}
\usepackage{float}
\usepackage{dcolumn}
\usepackage[nolist,nohyperlinks]{acronym}
\usepackage{xspace}
\usepackage[abs]{overpic}
\usepackage{pict2e}
\usepackage{enumitem}
\usepackage[usenames,dvipsnames]{xcolor}
\usepackage[utf8]{inputenc}
\usepackage{acronym}
\usepackage{gensymb}
\usepackage{cleveref}
\usepackage[normalem]{ulem}
\usepackage{lineno}
\usepackage{mathrsfs}
\usepackage{array}

\usepackage[caption=false]{subfig} 

\DeclareMathAlphabet{\mathcalstd}{OMS}{cmsy}{m}{n}
\DeclareMathAlphabet{\mathpzc}{OT1}{pzc}{m}{it}

\renewcommand{\today}{\number\day\space\ifcase\month\or
  January\or February\or March\or April\or May\or June\or
  July\or August\or September\or October\or November\or December\fi
  \space\number\year}

\newcommand{\TheEvent}{GW150914}

\begin{document}


\title{Effects of waveform model systematics on the interpretation of \TheEvent{}}
\author{B.\,P.~Abbott \emph{et al.}}
\email{Full author list given at the end of the article}
\noaffiliation{}


\begin{abstract}
  Parameter estimates of \TheEvent{} were obtained using Bayesian inference, based on three
  semi-analytic waveform models for binary black hole coalescences.
  These waveform models differ from each other in
  their treatment of black hole spins, and all three
  models make some simplifying assumptions, notably to neglect
  sub-dominant waveform harmonic modes and orbital eccentricity.
  Furthermore, while the models are calibrated to agree with waveforms
  obtained by full numerical solutions of Einstein's equations, any
  such calibration is accurate only to some non-zero tolerance and is
  limited by the accuracy of the underlying phenomenology, availability,
  quality, and parameter-space coverage of numerical simulations.
  This paper complements the original analyses of \TheEvent{} with an investigation of
  the effects of possible systematic errors in the waveform models on estimates of its
  source parameters.
  To test for systematic errors we repeat the original Bayesian analyses on mock
  signals from numerical simulations of a series of binary configurations with
  parameters similar to those found for \TheEvent{}.
  Overall, we find no evidence for a systematic bias relative to the statistical error
  of the original  parameter recovery of \TheEvent{} due to modeling approximations or
  modeling inaccuracies.  However, parameter biases are found to occur for some
  configurations disfavored by the data of \TheEvent{}: for binaries inclined
  edge-on to the detector over a small range of choices of polarization angles, and
  also for eccentricities greater than $\sim$0.05. For signals with higher
  signal-to-noise ratio than \TheEvent{}, or in other regions of the binary parameter
  space (lower masses, larger mass ratios, or higher spins), we expect that systematic
  errors in current waveform models may impact gravitational-wave measurements, making
  more accurate models desirable for future observations.
\end{abstract}


\maketitle

\acrodef{aLIGO}{Advanced Laser Interferometer Gravitational wave Observatory}
\acrodef{AdV}{Advanced Virgo}
\acrodef{BBH}[BBH]{binary black hole}
\acrodef{BNS}[BNS]{binary neutron star}
\acrodef{NS}[NS]{neutron star}
\acrodef{BHNS}[BHNS]{black hole--neutron star binaries}
\acrodef{PBH}[PBH]{primordial black hole binaries}
\acrodef{SNR}[SNR]{signal-to-noise ratio}
\acrodef{LIGO}[LIGO]{Laser Interferometer Gravitational-Wave Observatory}
\acrodef{LHO}[LHO]{LIGO Hanford}
\acrodef{LLO}[LLO]{LIGO Livingston}
\acrodef{LSC}[LSC]{LIGO Scientific Collaboration}
\acrodef{CBC}[CBC]{compact binary coalescence}
\acrodef{GW}[GW]{gravitational wave}
\acrodef{FAR}[FAR]{false alarm rate}
\acrodef{FAP}[FAP]{false alarm probability}
\acrodef{IFO}[IFO]{interferometer}
\acrodef{BH}[BH]{black hole}
\acrodef{GR}[GR]{general relativity}
\acrodef{NR}[NR]{numerical relativity}
\acrodef{PN}[PN]{post-Newtonian}
\acrodef{EOB}[EOB]{effective-one-body}
\acrodef{PSD}[PSD]{power spectral density}
\acrodef{PDF}[PDF]{probability density function}
\acrodef{PE}[PE]{parameter estimation}
\acrodef{LAL}[LAL]{LIGO Algorithm Library}
\acrodef{IMR}[IMR]{inspiral-merger-ringdown}
\acrodef{BSSN}[BSSN]{Baumgarte-Shapiro-Shibata-Nakamura }
\acrodef{ROM}[ROM]{reduced-order-model}
\acrodef{DPF}[DPF]{dominant polarisation frame}

\newcommand{\PN}[0]{\ac{PN}\xspace}
\newcommand{\BBH}[0]{\ac{BBH}\xspace}
\newcommand{\NR}[0]{\ac{NR}\xspace}
\newcommand{\GW}[0]{\ac{GW}\xspace}
\newcommand{\SNR}[0]{\ac{SNR}\xspace}
\newcommand{\aLIGO}[0]{\ac{aLIGO}\xspace}
\newcommand{\AdV}[0]{\ac{AdV}\xspace}
\newcommand{\BH}[0]{\ac{BH}\xspace}
\newcommand{\PE}[0]{\ac{PE}\xspace}
\newcommand{\IMR}[0]{\ac{IMR}\xspace}
\newcommand{\PDF}[0]{\ac{PDF}\xspace}
\newcommand{\GR}[0]{\ac{GR}\xspace}
\newcommand{\PSD}[0]{\ac{PSD}\xspace}


\section{Introduction}
\label{sec:introduction}

We recently reported the first direct observation of a \GW signal by the \aLIGO,
from the merger of two black holes, \TheEvent{}~\cite{Abbott:2016blz}.
The merger occurred at a distance of $\sim$410\,Mpc and the black holes were estimated to have masses of
$\sim$36~$M_\odot$ and $\sim$29~$M_\odot$, with spins poorly constrained to be each $<$0.7 of their maximum possible
values; we discuss the full properties of the
source in detail in Refs.~\cite{PhysRevLett.116.241102,Abbott:2016izl}, subsequently refined in~\cite{TheLIGOScientific:2016pea}. These parameter
estimates relied on three semi-analytic models of \BBH \GW signals~\cite{Taracchini:2013rva,Purrer:2015tud,Hannam:2013oca,Khan:2015jqa,Pan:2013rra,Babak:2016tgq}.
In this paper we investigate systematic parameter errors that may have resulted from the approximations or physical
infidelities of these waveform models, by repeating the analysis of Ref.~\cite{PhysRevLett.116.241102} using a set of
\NR waveforms from configurations similar to those found for \TheEvent{}.

The dynamics of two black holes as they follow a non-eccentric orbit, spiral towards each other and merge, are determined by the
\BH masses $m_1$ and $m_2$, and the \BH spin angular momenta $\mathbf{S}_1$ and $\mathbf{S}_2$. The resulting \GW signal
can be decomposed into spin-weighted spherical harmonics, and from these one can calculate the signal one would
observe for any orientation of the binary with respect to our detectors. For binary systems where the detectors are
sensitive to the signal from only the last few orbits before merger (such as \TheEvent{}), we can calculate the
theoretical signal from \NR solutions of the full nonlinear Einstein equations (see e.g.~\cite{alcubierrebook,baumgartebook}).
However, since the computational cost of NR simulations is substantial, in practice Refs.~\cite{PhysRevLett.116.241102,Abbott:2016izl,TheLIGOScientific:2016pea} utilized semi-analytic models that can be evaluated millions of times to measure the source properties.

The models used in the analysis of \TheEvent{}, non-precessing
EOBNR~\cite{Taracchini:2013rva,Purrer:2015tud},
IMRPhenom~\cite{Hannam:2013oca}, and precessing
EOBNR~\cite{Pan:2013rra,Taracchini:2013rva,Babak:2016tgq}, estimate the
dominant \GW harmonics for a range of \BBH systems, incorporating
information from \PN theory for the inspiral, and the \ac{EOB} approach for the entire coalescing process, and
inputs from \NR simulations~\cite{Mroue:2013xna,Husa:2015iqa}, which
provide a fully general-relativistic prediction of the \GW signal from
the last orbits and merger.
Non-precessing EOBNR represents signals from binaries where the \BH-spins
are aligned (or anti-aligned) with the direction of the binary's orbital angular momentum\footnote{In this work, we always refer to the Newtonian angular momentum, often denoted $L_N$ in the technical literature~\cite{blanchet:2013haa}.}, $\mathbf{\hat{L}}$. In such systems the orbital plane remains fixed (i.e., $\hat{\mathbf{L}}=\mathrm{const.}$) and the binary is parameterized only by each \BH mass and each dimensionless spin-projection onto $\hat{\mathbf{L}}$,
\begin{equation}\label{eq:chi_L}
\chi_{iL}\equiv \frac{c\mathbf S_i\cdot\hat{\mathbf{L}}}{Gm_i^2},
\end{equation}
where $i=1,2$ labels the two black holes, $c$ denotes the speed of light and $G$ Newton's constant.

For binaries with generic \BH-spin orientations, the orbital plane no longer remains fixed (i.e., $\hat{\mathbf{L}} \neq \mathrm{const.}$), and such binaries
exhibit precession caused by the spin components orthogonal
to $\hat{\mathbf{L}}$~\cite{apostolatos:1994, kidder:1995zr}.
Depending on the orientation of the orbital plane relative to an observer at a reference epoch (time or frequency), the inclination of the binary is defined as
\begin{equation}
\label{eq:inclination}
\iota := \arccos{(\hat{\mathbf L} \cdot \hat{\mathbf N})},
\end{equation}
where $\hat{\mathbf N}$ denotes the direction of the line-of-sight from the binary to the observer,
precession-induced modulations are observed in the \GW signal.
The waveforms from such binaries are modeled in precessing EOBNR and precessing IMRPhenom.
The precessing IMRPhenom waveform model incorporates precession effects through a single
precession spin parameter $\chi_p$~\cite{schmidt:2014iyl} and one spin-direction within the instantaneous orbital plane.
These parameters
are designed to capture the dominant precession-effects, which are described through approximate PN results.
Precessing EOBNR utilizes an effective-one-body Hamiltonian and radiation-reaction force that
includes the full six spin degrees of freedom, which are evolved using
Hamilton's equations of motion.  Both precessing models ---IMRPhenom and EOBNR--- are calibrated only against
non-precessing numerical simulations, although both models were compared with precessing numerical simulations~\cite{Pan:2013rra,Taracchini:2013rva,Babak:2016tgq,boheetalinprep}.  While inclusion of the complete spin-degrees of freedom
in the precessing EOBNR models can be advantageous~\cite{Babak:2016tgq}, precessing EOBNR suffers the practical limitation of high computational cost.
For that reason complete \PE results using precessing EOBNR for \TheEvent{} were published separately~\cite{Abbott:2016izl}, and \PE for the two other \BBH events reported during the first \aLIGO observing run, LVT151012 and GW151226~\cite{Abbott:2016nmj,TheLIGOScientific:2016pea}, are presently only available based on non-precessing EOBNR and precessing IMRPhenom.
We also restrict this study to non-precessing EOBNR and precessing IMRPhenom but include non-precessing IMRPhenom in select studies.

Apart from the treatment of spin (aligned-spin, effective precession-spin, full-spin), all
waveform models discussed include errors due to the limited number of \NR calibration waveforms, the inclusion
of only the strongest spherical-harmonic modes, and
 the assumption of a non-eccentric inspiral.
Our previous analyses~\cite{Abbott:2016nmj,TheLIGOScientific:2016pea,Abbott:2016izl}
indicated that \TheEvent{} is well within the parameter region over which the models were calibrated, and most
likely oriented such that any precession has a weak effect.  The goal of the present study is to ensure
that our results are not biased by the limitations of the waveform models. We achieve this by performing the parameter recovery on a set of
\NR waveforms, which are complete calculations of the \GW signal that include the full harmonic content of the signal,
limited only by small numerical inaccuracies that we show are insignificant in the context of this analysis.

Our analysis is based on injections of numerical waveforms into simulated or actual
detector data, i.e. we add gravitational waveforms from \NR to the data-stream, and analyze
the modified data. We consider a set of \NR waveforms as mock \GW signals and extract
the source properties with the same methods that were used in analyzing \TheEvent{}
\cite{PhysRevLett.116.241102}, with two main differences: (1) by injecting \ac{NR}
waveforms as mock signals, we know the exact parameters of the waveform being
analyzed. This knowledge allows us to compare the \PDF obtained by our
blind analysis with the simulated parameters of the source; (2) in order to
assess the systematic errors independently of the statistical noise
fluctuations, we use the estimated \PSD from actual \aLIGO data
around the time of \TheEvent{} as the appropriate weighting in the inner
product between the \NR signal and model waveforms.
We use injections into ``zero noise'' where the data is composed of zeros plus the mock signal. This makes our analysis independent of a concrete (random) noise-realization so that our results can be interpreted as an average over many Gaussian noise realizations.
The detector noise curve enters only in the power spectral density which impacts the likelihood-function Eq.~(\ref{eqn:likelihood_ratio}) through the noise-weighted inner product, Eq.~(\ref{eq:inner-product}). This allows us to properly include the characteristics of \aLIGO's noise.

While systematic errors could be assessed from the computation of the fitting factor~\cite{apostolatos:1995}, Bayesian parameter estimation has several advantages: (i) it provides information about the statistical error at the \SNR that the signal is seen at in the detector network; (ii) it performs a detailed and robust sampling of the vicinity of the best fit parameters; (iii) it properly includes the response of a detector network; (iv) it replicates the setup of the parameter estimation analyses used for GW150914 and therefore enables immediate comparisons.

\begin{table}
  \centering
   \begin{tabular}{p{0.4\columnwidth} >{\centering\arraybackslash}p{0.28\columnwidth} >{\centering\arraybackslash}p{0.28\columnwidth}}
\hline
\hline
   & injected value & \TheEvent{} \\
   \hline
   signal-to-noise-ratio $\rho$ &  25 & 23.7  \\
   $f_\mathrm{ref}$ & 30~Hz &20~Hz \\
   detector-frame total mass&  $74.10$~$M_\odot$ & $70.6^{+4.6 \pm 0.5}_{-4.5 \pm 1.3}$~$M_\odot$   \\
   inclination $\iota$ & $162.55^\circ$ & -- \\
   polarisation angle $\psi$ & $81.87^\circ$ & -- \\
   right ascension $\alpha$ & $07^h 26^m 50^s$ & -- \\
   declination $\delta$ & $-72.28^\circ$ & -- \\
\hline
\hline
\end{tabular}
  \caption{Fiducial parameter values chosen for all analyses unless stated
  otherwise and the corresponding estimates for \TheEvent{}.
  $\iota$ denotes the inclination at the reference frequency $f_\mathrm{ref}$, $\psi$ the
  polarization angle and $(\alpha, \delta)$ the location of the source in the sky
  (see Sec.~\ref{sub:infrastructure_for_injecting_nr_waveforms_into_ligo_data} for details).
  The total mass and the four angle parameters are chosen from within the $90\%$ credible intervals
  obtained in the Bayesian analysis presented in Ref.~\cite{PhysRevLett.116.241102}, where the
  polarization angle was found to be unconstrained, the inclination strongly disfavored to be
  misaligned with the line-of-sight and results for the sky location are depicted in Fig. 4
  therein.
  }
  \label{tab:fiducial}
\end{table}

We test a variety of binary configurations in the vicinity of \TheEvent{}'s
parameters.
In particular, for several analyses we choose \ac{NR} configurations\footnote{Due to the time required to produce NR simulations, these were initiated shortly after the detection of \TheEvent{} when final parameter estimates were not yet available.  Thus, our fiducial values differ slightly from the final parameter estimates.} at \emph{fiducial}
parameter values consistent with those found for \TheEvent{}~\cite{PhysRevLett.116.241102}.
Those fiducial parameter values are listed in Table~\ref{tab:fiducial} in comparison to
the parameter estimates for \TheEvent{}~\cite{PhysRevLett.116.241102}.

Throughout this paper we will be using two effective spin parameters to represent spin information: First, an effective inspiral spin, $\chi_\mathrm{eff}$,
defined by~\cite{ajith:2009bn, santamaria:2010yb}
\begin{equation}
\label{eq:chieff}
\chi_\mathrm{eff}:=\frac{m_1 \chi_{1L} + m_2\chi_{2L}}{m_1 + m_2},
\end{equation}
where $\chi_{iL}$ is defined in Eq.~(\ref{eq:chi_L}), and, second, an effective
precession spin, $\chi_p$, a single spin parameter that captures the dominant spin
contribution that drives precession during the inspiral (cf. Eq.~(3.4)
in Ref.~\cite{schmidt:2014iyl}).

We first address the recovery of the binary parameters for aligned-spin systems, and then
for precessing ones. Then we study the effect of different polarization angles and inclinations
and address the question of the influence of higher harmonics.
Since the semi-analytic waveform models only model a quasi-spherical orbital evolution,
we also investigate any biases related to residual eccentricity. Finally, we investigate
the effects of non-stationary detector noise and numerical errors.

Our study can only determine whether our waveform models would incur a bias in the
parameters we have measured, but cannot tell us whether further information could be
extracted from the signal, e.g., additional spin information beyond the effective precession
parameter in IMRPhenom, or eccentricity. The parameter-estimation analysis of \TheEvent{} with
precessing EOBNR provides slightly stronger constraints
on the \BH spins, but limited additional spin information~\cite{Abbott:2016izl}. No studies have yet been performed to
estimate the eccentricity of \TheEvent{}, although preliminary investigations have bounded the eccentricity to be $\leq 0.1$~\cite{PhysRevLett.116.241102}.

The strategy pursued here ---inject known synthetic (\NR) signals, recover with waveform models--- joins a complementary study that analyses the \TheEvent-data based directly on numerical waveforms~\cite{Abbott:2016apu} instead of making use of semi-analytic waveform models.
In another study~\cite{Jani:2016wkt}, additional \NR simulations were performed for
parameters similar to \TheEvent{}. Those, along with other simulations from~\cite{Jani:2016wkt}, were compared with the reconstructed signal of \TheEvent{} from unmodelled searches, and
provided an independent check on source parameters~\cite{TheLIGOScientific:2016uux}.

The remainder of this paper is organized as follows: In Sec.~\ref{sec:methodology} we describe
the \ac{NR} waveforms we us as mock GW signals, and summarize the Bayesian \PE algorithm,
the \ac{NR} injection framework and the waveform models, which our NR injections are compared against. In Sec.~\ref{sec:results}
we present the results of our various analyses. Our main result is that the different waveform models used in the analysis
of \TheEvent{} did \emph{not} induce any significant systematic errors in our measurements of this source.
We conclude in Sec.~\ref{sec:discussion}.


\section{Methodology}
\label{sec:methodology}

\subsection{Waveform models}
\label{sub:waveform_models}

In this section we briefly describe the waveform models used to
measure the properties of \TheEvent{}.  We summarize the assumptions
and approximations behind each model, and highlight their domains of
validity in the \BBH parameter space.  All models aim to represent
non-eccentric \BBH inspirals. Such binaries are governed by
eight source-intrinsic degrees of freedom: The two black hole masses
$m_{1,2}$, and the two dimensionless black hole spin-vectors
$\boldsymbol{\chi}_{1,2}$.  The waveform measured by a \GW detector on
Earth also depends on extrinsic parameters that describe the relative
orientation of the source to the detector, bringing the total number
of degrees of freedom in parameter estimation to 15.
Owing to the scale-invariance of vacuum \GR, the total mass scales out
of the problem, and a seven-dimensional intrinsic parameter space remains, which
is spanned by the mass-ratio $q \equiv m_2/m_1 \leq 1$, and the
spin-vectors.  The spatial emission pattern of the emitted \ac{GW} is
captured through spherical harmonic modes $(\ell, m)$ defined in a suitable coordinate system. All models
used here represent only the dominant quadrupolar gravitational-wave
emission, corresponding approximately to the $(\ell, m)=(2,\pm 2)$
spherical-harmonic modes.

Two main approaches to construct analytical \ac{IMR} waveform models have been developed in recent years:
the \ac{EOB} formalism and the phenomenological framework.

The \ac{EOB} approach to modeling the coalescence of compact-object binaries was first introduced in Refs.~\cite{Buonanno:1998gg, Buonanno:2000ef} as a way to
extend the \ac{PN} results of the inspiral
to the strong-field regime, and model semi-analytically the merger and ringdown stages.
In this approach, the conservative \ac{PN} dynamics of a pair of \acp{BH} is mapped to the motion of a test particle moving in a deformed Kerr spacetime,
where the deformation is proportional to the symmetric mass ratio,
$\nu = m_1 m_2/(m_1 + m_2)^2$, of the binary.
Prescriptions for resumming \ac{PN} formulas of the waveform modes are used to construct inspiral-plunge GW signals~\cite{Damour:2007xr,Damour:2008gu,Pan:2010hz}.
To improve agreement with \ac{NR} waveforms, high-order, yet unknown PN terms are inserted in the \ac{EOB} Hamiltonian and tuned to \ac{NR} simulations. Also, additional terms are included in the waveform to improve the behavior during plunge and merger (e.g., non-quasi-circular corrections) and to optimize agreement with \ac{NR} waveforms.
The ringdown signal is modeled as a linear combination of quasi-normal modes~\cite{Berti:2005ys, London:2014cma} of the remnant \ac{BH}.
A number of EOB models were developed for non-spinning~(e.g.~\cite{dn2008,pan:2011gk}), aligned-spin (e.g.~\cite{taracchini:2012ig,Damour:2014sva}), and generic spin-orientations (i.e., precessing systems, e.g.~\cite{Pan:2013rra}). They differ mainly by the underlying \ac{PN} resummation and \ac{NR} waveforms used to calibrate them. All current \ac{EOB} models rely on \ac{NR} simulations~\cite{dn2008,boyle:2008ge,Lovelace:2011nu,Damour:2011fu,Buchman:2012dw,Mroue:2013xna,Nagar:2015xqa} to achieve a reliable representation of the late inspiral, merger and ringdown portion of the BBH waveforms.

The present study is based on the \ac{EOB} model of Ref.~\cite{Taracchini:2013rva} for aligned-spin \ac{BBH}, which we shall refer to as non-precessing EOBNR.\footnote{The technical name of this model in LALSuite is \texttt{SEOBNRv2}.}
This model was calibrated to \ac{NR} waveforms with mass ratios between 1 and 1/8, and spins $-0.95 \leq \chi_{iL} \leq 0.98$ for $q=1$, as well as $-0.5 \leq \chi_{iL} \leq 0.5$ for $q\neq 1$.
This waveform model can be evaluated for arbitrary mass ratios and spin magnitudes. While the model can be evaluated outside its calibration region, its accuracy there is less certain than within the calibration region.
This model was extensively validated against independent \ac{NR} simulations~\cite{hinder:2013oqa,Taracchini:2013rva,Szilagyi:2015rwa,Kumar:2015tha,Kumar:2016dhh,Babak:2016tgq}, but the time integration of the EOB equations makes its evaluation computationally expensive. For the comprehensive \ac{PE} studies presented here, we therefore employ a frequency-domain \ac{ROM} ~\cite{Purrer:2015tud,Purrer:2014fza} of the non-precessing EOBNR.\footnote{In LALSuite,  this model is denoted by \texttt{SEOBNRv2\_ROM\_DoubleSpin}.} Recently, Ref.~\cite{Devine:2016ovp} introduced several optimizations that have significantly reduced the waveform generation time of the time-domain implementation, although the reduced-order-model remains significantly faster.

The generic-spin precessing time-domain EOBNR model\footnote{In LALSuite this model is denoted by \texttt{SEOBNRv3}.}~\cite{Pan:2013rra,Babak:2016tgq}
derives its orbital- and spin-dynamics from the
\ac{EOB} Hamiltonian of Refs.~\cite{Barausse:2009xi,Barausse:2011ys},
which incorporates all six spin-degrees of freedom.
The current precessing EOBNR model directly uses
calibration parameters from the non-precessing EOBNR model, without
renewed calibration on precessing \ac{NR} simulations.
The ringdown signal is generated in the frame aligned with the spin of the remnant
BH as a superposition of quasi-normal modes. The full \ac{IMR} waveform (as
seen in the inertial frame of an observer) is obtained by a
time-dependent rotation of the waveform modes in a suitable non-inertial
frame, i.e. the precessing-frame~\cite{boyle:2011gg}, according to
the motion of the Newtonian angular momentum, and by a constant
rotation of the ringdown.  An extensive
comparison~\cite{Babak:2016tgq} of the precessing EOBNR model to
precessing \ac{NR} simulations with mass ratios $1 \geq q \geq 1/5$
and dimensionless spin magnitudes $cS_i/(Gm_i^2) \leq 0.5$, finds remarkable agreement
between them.  Extensive code optimizations~\cite{Devine:2016ovp} have
also been performed on this model, but unfortunately, its computational
cost still prohibits its use in the present study.

The phenomenological approach exclusively focuses on the gravitational waveform without providing a description of the binary dynamics. It is aimed at constructing a closed-form expression of the GW signal in the frequency domain for computational efficiency in GW data analysis.  Phenomenological models were first introduced in Refs.~\cite{ajith:2007qp, Pan:2007nw} to describe the \ac{IMR} waveforms of non-spinning binaries.

IMRPhenom models are built on a phenomenological ansatz for the frequency-domain amplitude and phase of the \ac{IMR} signal, commonly an analytic extension of an inspiral description, e.g., from \ac{PN} theory, through merger and ringdown.  The functional form of such an extension is chosen based on inspection of \ac{NR} simulations, and free coefficients in the ansatz are calibrated against the \ac{NR} waveforms.

Phenomenological models of the dominant $(\ell,m)=(2,\pm 2)$ multipolar modes of the waveform were constructed for non-spinning and aligned-spin \ac{BBH}~\cite{ajith:2007qp,Pan:2007nw,ajith:2009bn,santamaria:2010yb,Husa:2015iqa,Khan:2015jqa}.
We refer to the most recent aligned-spin phenomenological waveform model as non-precessing IMRPhenom~\cite{Husa:2015iqa,Khan:2015jqa}.\footnote{In LALSuite this model is denoted as \texttt{IMRPhenomD}.}
Non-precessing IMRPhenom was calibrated against \ac{NR} waveforms with mass ratios between 1 and 1:18 with $-0.95 \leq \chi_{iL} \leq 0.98$ for $q=1$, and $-0.85 \leq \chi_{iL} \leq 0.85$ for $q \neq 1$.
Non-precessing IMRPhenom is calibrated with respect to an effective spin
parameter (cf. Eq. (\ref{eq:chieff})), although information from both \BH spins enters through the underlying \PN phasing and an alternative spin parameterization in the ringdown.
Comparisons with independent \ac{NR} waveforms indicate~\cite{Kumar:2016dhh} that non-precessing IMRPhenom is an accurate waveform model in the region of parameter space of relevance to \TheEvent.

For precessing binaries with generic \ac{BH} spin-directions, a direct phenomenological ansatz for the now seven-dimensional physical parameter space is impractical due to a lack of \ac{NR} simulations and the complicated precession-induced modulations in the waveforms.
This problem is addressed by rotating the waveform-modes of \emph{aligned-spin} \ac{BBH}
systems into precessing waveform-modes via a time-dependent rotation which describes the
motion of the orbital plane of the precessing \ac{BBH} under
consideration~\cite{Hannam:2013oca, schmidt:2010it, schmidt:2012rh}.
In this transformation, $\chi_{iL}$ are preserved~\cite{schmidt:2012rh}.
The most recent precessing IMRPhenom
model~\cite{Hannam:2013oca}\footnote{In LALSuite this model is denoted
  by \texttt{IMRPhenomPv2}.} used here makes several additional
approximations, in order to arrive at a closed-form frequency domain
expression: First, the four in-plane spin components are mapped into
one effective in-plane precession spin,
$\chi_p$~\cite{schmidt:2014iyl}, and one azimuthal
orientation.\footnote{As a result of this effective spin reduction,
the waveform seen by any given detector depends on 13 independent
parameters instead of the actual 15 degrees of freedom present in a
full description of the source.} Secondly, the precession of the
orbital plane is described with orbit-averaged \ac{PN} equations for
single-spin \ac{BBH} systems. Finally, the stationary-phase
approximation is used to derive the frequency domain expressions for
the ($\ell=2$) multipolar modes of the waveform in the inertial frame
of the observer. Both the \ac{PN} description and the stationary-phase
approximation are carried through merger. The spin of the
final black hole is also modified due to the in-plane spins.
We stress that the precessing sector of the precessing IMRPhenom model was not calibrated to \ac{NR} simulations, although comparison with precessing \ac{NR} simulations indicate promising agreement~\cite{boheetalinprep}.

The non-precessing models represent the dominant $(\ell=2,m=\pm 2)$ harmonics of the \GW signal to within a mismatch accuracy of
$\sim$1\% across their respective calibration regions~\mbox{\cite{Taracchini:2013rva,Purrer:2015tud,Khan:2015jqa}}
(mismatches are defined in Sec.~\ref{sub:infrastructure_for_injecting_nr_waveforms_into_ligo_data}), and to within $\sim$0.5\% in the
region of parameter space near \TheEvent{}; see Ref.~\cite{Kumar:2016dhh} for detailed comparisons of both the non-precessing EOBNR
and IMRPhenom models. Formally, two models are considered indistinguishable if the mismatch $1- \mathcal{O}$ between them satisfies
$1 - \mathcal{O} < 1/(2 \rho^2)$~\cite{Flanagan:1997kp,lindblom:2008cm,McWilliams:2010eq}, where $\rho$ is the \SNR. For signals with $\rho=25$, this corresponds to a mismatch error
better than 0.08\%, a bound not reached by the waveform models.
However, the indistinguishability criterion is only a sufficient condition, and its violation does not necessarily lead to systematic measurement biases in practice.

The waveform differences may be orthogonal to the physical signal manifold~\cite{Flanagan:1997kp}; we are dealing with a multidimensional parameter space and waveform differences may be distributed over many parameters; and the waveform errors for individual cases may be oscillatory and average out in the median while increasing the spread of marginal posterior distributions. This has been illustrated in Ref.~\cite{Littenberg:2012uj}, and is also exemplified by the results of the present study.

\subsection{Numerical relativity waveforms}
\label{sec:numerical_relativity_waveforms}

We use \ac{NR} waveforms produced by two independent codes, which use
completely different analytical and numerical methods, the Spectral
Einstein Code
({\tt SpEC})~\cite{SpECwebsite,Scheel:2006gg,Szilagyi:2009qz,Buchman:2012dw}
and the bifunctional adaptive mesh (BAM)
code~\cite{PhysRevD.77.024027,husa2008}.

The BAM code solves the Einstein evolution equations using the
\ac{BSSN}~\cite{Shibata:1995we,PhysRevD.59.024007} formulation of the
3+1-decomposed Einstein field equations.
The BSSN equations are integrated with a fourth-order finite-difference
Runge-Kutta time integrator, with Berger-Oliger time-stepping, along with
sixth-order accurate spatial finite differencing, based on the method-of-lines
for spatial derivatives.
The $\chi$-variation~\cite{campanelli:2005dd} of the moving-puncture method~\cite{baker:2005vv, campanelli:2005dd}
is used, where a new conformal
factor defined is $\chi_{g} = \psi^{-4}$, which is finite at the puncture.
The lapse and shift gauge functions are evolved using the 1+log slicing
condition and the Gamma-driver shift condition respectively.
Conformally flat puncture initial data \cite{Cook:1989fb,Brandt:1997tf,Bowen:1980yu}
are calculated using the pseudospectral elliptic solver described in \cite{Ansorg:2004ds}.
\BBH simulations were produced by the BAM code \cite{PhysRevD.77.024027,husa2008,Abbott:2016apu} with approximately random initial configurations within the 99\% credible region inferred for \TheEvent{} \cite{Abbott:2016blz}.
The current study utilizes three of these simulations, the parameters of which are detailed in Table \ref{tbl:NRparams},
which include all harmonic multipoles up to $(\ell=5)$.

The Spectral Einstein Code ({\tt SpEC})~\cite{SpECwebsite,Scheel:2006gg,Szilagyi:2009qz,Buchman:2012dw}
is a multi-domain, pseudospectral collocation code primarily used for
the simulation of compact object binary spacetimes. It is capable of
efficiently solving a wide array of hyperbolic and elliptic partial
differential equations with high accuracy. Conformally curved initial
data~\cite{Lovelace:2008tw} is constructed in the extended-conformal-thin-sandwich (XCTS)
formalism~\cite{Pfeiffer:2002iy}, using the {\tt SpEC} pseudo-spectral
elliptic solver~\cite{Pfeiffer:2002wt}. For evolution, {\tt SpEC} uses
the first order formulation~\cite{Lindblom2006} of the generalized
harmonic form of Einstein's equations~\cite{Pretorius2005a, Friedrich1985} in the damped harmonic
gauge~\cite{Lindblom:2009tu}.  Adaptive mesh refinement ensures
to achieve high accuracy and
efficiency~\cite{Szilagyi:2014fna}. Dynamical excision is used to
handle black hole
singularities~\cite{Hemberger:2012jz,Scheel:2014ina}. {\tt SpEC} has
been successfully employed to study many aspects of modelling compact
object binaries (see e.g
~\cite{Szilagyi:2015rwa,Barkett:2015wia,Lovelace:2014twa,Mroue:2013xna,Buchman:2012dw}).
In this study we use waveforms from the SXS public catalog~\cite{Mroue:2013xna,SXSCatalog},
which has seen recent additions of 90 aligned-spin waveforms~\cite{Chu:2015kft}, as well as
new simulations targeted by \TheEvent{}~\cite{Blackman:2017dfb}.
All the SXS simulations have $>24$ GW cycles
and start below 30~Hz at $74$~$M_{\odot}$.The SXS waveforms include all harmonics
up to and including $(\ell=8)$. The waveforms are extracted on a series of concentric
coordinate spheres of various radii and the data is then extrapolated to null
infinity with polynomial extrapolation~\cite{Boyle:2009vi}. For precessing configurations, the drift of the
center of mass due to residual initial linear momentum is corrected using the method described in~\cite{Boyle:2015nqa}
to avoid any spurious mixing of \GW modes.

Both codes are described in more detail in Ref.~\cite{hinder:2013oqa}. Their results (along
with those from three other codes) were found to be sufficiently accurate and consistent
for \aLIGO observations of equal-mass nonspinning binaries up to
an \SNR of $\sim$25~\cite{physrevd.79.084025}, which is similar to the expected configuration of
\TheEvent{}. A recent study comparing {\tt SpEC} and another moving-punctures code, LazEv~\cite{campanelli:2005dd},
also found excellent agreement (with a mismatch $\lesssim 1\times10^{-3}$ for \aLIGO design sensitivity) between waveforms for an aligned-spin binary with parameters consistent with \TheEvent{}~\cite{SpECvsLazEv}.
Waveforms from the BAM code and \texttt{SpEC} \emph{not} used as mock signals in this study were used in the construction of the EOBNR and IMRPhenom models.

\begin{table*}
  \begin{ruledtabular}
  \begin{tabular}{lccccccccc}
    ID & $q$ & $\boldsymbol{\chi}_{1}$ & $\boldsymbol{\chi}_{2}$ & $\chi_{\rm eff}$ & $\chi_{p}$
& $M\Omega$ & $N_{\rm orbits}$ & $e$ & $1 - \mathcal{O}_{\rm res}$\\
\hline
\texttt{SXS:BBH:0307} & 0.813 &  (0,0,0.32) & (0,0,-0.58) & -0.0839 & 0.0 &   0.01972 & 12.6 & $4\times10^{-4}$  & $2\times 10^{-5}$\\
\texttt{SXS:BBH:0308} & 0.813 & (0.0943,0.0564,0.3224)  & (0.2663,0.2134,-0.5761) & -0.0842 & 0.2629 & 0.019729 & 12.5 & $5\times10^{-4}$ & $1.4\times 10^{-4}$ \\
\texttt{CFUIB0020} & 0.833 & (-0.2594, -0.5393, -0.0458) & (-0.0276, -0.2194, 0.3622) & 0.1396 & 0.5985 & 0.0276 & 6.6 & $3.6\times 10^{-3}$ & n/a \\
\texttt{CFUIB0012} & 0.833 & (-0.1057, 0.2362, 0.1519) & (0.1269, -0.5130, 0.4139) & 0.2710 & 0.4291 & 0.0268 & 7.4 & $7.7\times 10^{-3}$ & n/a \\
\texttt{CFUIB0029} & 0.833 & (-0.2800, -0.2896, -0.1334) & (0.3437, 0.2283, 0.0989) & -0.0278 & 0.4028 & 0.0268 & 6.5 & $3.7\times 10^{-3}$ &  $7.3\times10^{-5}$ \\[6pt]
\texttt{SXS:BBH:0318} & 0.82 & (0,0,0.3300) & (0,0,-0.4399) & -0.0168 & 0.0 & 0.020539 &
14.1 & 0.0049 & $8.9\times 10^{-5}$ \\
\texttt{SXS:BBH:0319} & 0.82 & (0,0,0.3300) & (0,0,-0.4399) & -0.0168 & 0.0 & 0.020322 & 14.3 & 0.0091 & $1\times 10^{-4}$ \\
\texttt{SXS:BBH:0320} & 0.82 & (0,0,0.3300) & (0,0,-0.4399) & -0.0168 & 0.0 & 0.021330 & 13.5 & 0.013  & $3.1\times 10^{-5}$ \\
\texttt{SXS:BBH:0321} & 0.82 & (0,0,0.3299) & (0,0,-0.4399) & -0.0168 & 0.0 & 0.018717 & 15.0 & 0.029  & $6.5\times 10^{-5}$ \\
\texttt{SXS:BBH:0322} & 0.82 & (0,0,0.3301) & (0,0,-0.4399) & -0.0168 & 0.0 & 0.019363 & 15.0 & 0.038  & $4.9\times 10^{-4}$ \\
\texttt{SXS:BBH:0323} & 0.82 & (0,0,0.3300) & (0,0,-0.4400) & -0.0168 & 0.0 & 0.018392 & 14.6 & 0.070  & $2.2\times10^{-4}$\\
\texttt{SXS:BBH:0324} & 0.82 & (0,0,0.3299) & (0,0,-0.4400) & -0.0168 & 0.0 & 0.017351 & 13.0 & 0.13   & $8.7\times10^{-4}$\\
  \end{tabular}
  \caption{Primary simulations used in this study.  Given are
    mass-ratio $q$, dimensionless spin-vectors $\boldsymbol\chi_{1,2}$, and
    the corresponding effective aligned-spin parameter $\chi_{\rm
      eff}$ and precession-spin parameter $\chi_p$ of the precessing
    IMRPhenom model.  Spin-vectors are specified at the reference
    epoch where the orbital frequency $M\Omega$ takes the indicated
    value.  The last three columns give the number of orbits from the
    reference-epoch to merger, the orbital eccentricity at
    reference-epoch, and an approximate measure of the numerical
    truncation error respectively (see main text for details).}
\label{tbl:NRparams}
 \end{ruledtabular}
\end{table*}

\subsection{Gravitational waveform processing \& injection}
\label{sub:infrastructure_for_injecting_nr_waveforms_into_ligo_data}

Gravitational waveforms extracted from \NR simulations are commonly decomposed into time-dependent multipolar modes $h_{\ell m}(t)$ in a basis of
spherical harmonics ${}^{-2}Y_{\ell m}$ with spin weight $-2$.
The two \GW polarizations $h_+$ and $h_\times$ are given in terms of the $h_{\ell m}$-modes by
\begin{equation}
\label{eq:GW-polarizations}
h_+ - i h_\times = \sum_{\ell, m} h_{\ell m} {}^{-2}Y_{\ell m}.
\end{equation}
The numerical simulations provide the modes $h_{\ell m}$
evaluated in the coordinates of the numerical simulation (\NR-frame) sampled
at times determined by the numerical simulation.

We remove initial spurious radiation from the \NR data and align the
modes such that the peak of the waveform, defined as $h_\mathrm{peak}:=\max(\sum_{\ell, m} |h_{\ell, m}|^2)$,
occurs at the time $t=0$. We split each complex mode
into separate real-valued amplitude $A_{\ell m}$ and phase $\Phi_{\ell m}$ according to
\begin{equation}
\label{ }
h_{\ell m} = A_{\ell m} \exp(i \Phi_{\ell m}).
\end{equation}
Next, one-dimensional spline interpolants are constructed (separately on the
phase $\Phi_{\ell m}$ and the amplitude $A_{\ell m}$ of each mode
$h_{\ell m}$),
via a greedy algorithm to reduce the data size to a
fraction of its original value while guaranteeing reproducibility of
the original NR data to within a tolerance of $10^{-6}$~\cite{Galley:2016mvy}.

A common convention is to define a stationary
source frame which is aligned with the geometry of the binary at a certain reference epoch (detector-frame) as follows:
the $z$-axis is parallel to the orbital angular momentum direction, $\hat{L}$, and
the $x$-axis is parallel to the line $\hat n$ pointing from the less massive body $m_2$ to the more massive body $m_1$.
Two angles $\iota$ and $\phi$ then denote the
latitude and longitude respectively of the observer in that source
frame.
The (dimensionless) spins of the two bodies, $\boldsymbol{\chi}_i\equiv c\boldsymbol{S}_i/(Gm_i^2)$, $i=1,2$, are also expressed with respect to this stationary source frame:

\begin{align}
\label{eq:chix}
    \chi_{i\,x}& = \boldsymbol{\chi}_i^\mathrm{NR} \cdot \hat{\mathbf n},   \\
    \chi_{i\,y}& =  \boldsymbol{\chi}_i^\mathrm{NR} \cdot (\hat{\mathbf L} \times \hat{\mathbf n}), \\
\label{eq:chiz}
    \chi_{i\,z}& = \boldsymbol{\chi}_i^\mathrm{NR} \cdot \hat{\mathbf L}.
\end{align}

Given an emission direction determined by the angles $(\iota, \phi)$,
(in the detector-frame)
a total mass $M=m_1 + m_2$ and a desired sampling rate, the one-dimensional
spline interpolants are evaluated at the desired uniform time samples
to recover the numerical modes $h_{\ell m}(t)$.  The source frame
angles $(\iota, \phi)$ are transformed into the \ac{NR}-frame, and
Eq.~(\ref{eq:GW-polarizations}) is evaluated to compute the
\GW polarizations $h_+$ and $h_\times$.
This procedure is described
in~\cite{NRinjSchmidtHarryInPrep} and implemented in LAL~\cite{lal}.

The \GW data recorded by the \aLIGO detectors, $h_\mathrm{resp}$, are then obtained by projecting these \GW polarizations
onto each of the \aLIGO detectors via the antenna response functions $F_{+,\times}(\alpha, \delta, \psi)$ as follows:
\begin{align}
  \label{eq:detector_response}
  h_\mathrm{resp} = F_+(\alpha, \delta, \psi) \, h_+ + F_\times(\alpha, \delta, \psi) \, h_\times,
\end{align}
where $(\alpha, \delta)$ denote the right ascension and declination specifying the position of the \ac{GW} source in
the sky in an Earth-centered coordinate system, and $\psi$ is the polarization angle~\cite{anderson:2000yy, thorne.k:1987, injcosy}.

Most of our analyses focus on \NR injections into zero noise, but we also perform injections into calibrated strain data from the \aLIGO detectors LIGO-Hanford and LIGO-Livingston using tools
in the \texttt{PyCBC} software package~\cite{canton:2014ena, usman:2015kfa, pycbc, Biwer:2016ab}.

For all injections we choose a network \SNR of
25 and a low-frequency cut-off $f_{\rm low}=30$~Hz. The sampling rate is 16,384~Hz
and the waveforms are tapered at the start of the injection. We do not
apply a high-pass filter but add segment padding to remove any
high-pass corruption.

Our analyses commonly utilize a noise-weighted inner product between two waveforms
$a$ and $b$~\cite{Cutler:1994ys}:
\begin{equation}
\label{eq:inner-product}
  (a\,|\,b)= 4 \mathrm{Re} \int_{f_{\rm low}}^{f_{\rm high}} \frac{\tilde a(f) \tilde b^*(f)}{S_n(f)} df.
\end{equation}
Here $\tilde a(f)$ and $\tilde b(f)$ are the Fourier-transforms of the
real-valued
functions $a(t)$ and $b(t)$,
respectively and ${}^*$ denotes complex conjugation.
We use a high frequency cutoff of $f_{\rm high} = 2048$~Hz.
To estimate the median \ac{PSD} $S_n(f)$ used in this inner product,
we use 512s of \aLIGO data measured adjacent to the coalescence time of
\TheEvent{}.\footnote{The \ac{PSD} is generated from an 
  earlier calibration of the data but we have verified that it is
  accurate to within 1\% with a \ac{PSD} from the final calibration of the data.} 
The strain data stretches have previously been
calibrated such that the total uncertainty in the magnitude of the
recorded strain is less than 10\% and less than $10^\circ$ in phase
between 20~Hz and 1~kHz~\cite{Abbott:2016jsd}.

\subsection{Parameters of numerical simulations}

Table~\ref{tbl:NRparams} lists the parameters of the primary numerical simulations used in this study, whereas
Table~\ref{tbl:NRparams_additional} summarizes additional simulations that were employed for consistency checks in a wider region of parameter space.
The simulations shown in Table~\ref{tbl:NRparams} were specifically produced to follow-up \TheEvent{}.

For each simulation, parameters are given at the start of the useable numerical simulation, i.e. the reference epoch, indicated by a dimensionless (total mass invariant) orbital frequency
  $M\Omega$. This dimensionless orbital frequency $M\Omega$ translates into a
  gravitational-wave starting frequency of
\begin{equation}
f_{\rm GW} = \frac{M\Omega}{\pi}\left(GMc^3\right)^{-1},
\end{equation}
for the dominant $(2,2)$-harmonic.
For $M=74$~$M_\odot$,
  this corresponds to a frequency of $f_{\rm GW} = M\Omega\times 870$~Hz,
 so that dimensionless
  orbital frequencies $M\Omega$ of $0.027$ and $0.020$ translate to \GW frequencies of $23.5$~Hz,
  $17.4$~Hz, respectively.

Table~\ref{tbl:NRparams} specifies the two dimensionless spin vectors
$\boldsymbol{\chi}_i$ in
the LIGO-frame following Eqs.~(\ref{eq:chix})--(\ref{eq:chiz}).
The orbital eccentricity $e$ (at reference epoch) is estimated as follows: For the quasi-circular SXS simulations (\texttt{SXS:BBH:0307, SXS:BBH:0308} and Table~\ref{tbl:NRparams_additional}), a sinusoid is fitted to the time-derivative of the orbital frequency as detailed in~\cite{Buonanno:2010yk}.  For the CFUIB simulations, the eccentricity is measured with reference to an estimate of the non-eccentric
frequency evolution, which is found by fitting a fourth-order polynomial to the orbital frequency as in Ref.~\cite{Husa:2007rh}. More precise
estimates of the eccentricity can be made (for non-precessing signals) from the \GW signal, and these can be used to calculate initial
parameters for configurations with yet lower eccentricity, but we do not expect such low eccentricities to be necessary for this
study~\cite{Purrer:2012wy}; that expectation is supported by the results in Sec.~\ref{sub:eccentricity}.
Eccentric binaries exhibit a more complicated behavior of the orbital frequency.
For the eccentric simulations (\texttt{SXS:BBH:0318} to \texttt{SXS:BBH:0324}), therefore, we proceed as follows:  the \GW frequency is fitted according to Eqs.~(\ref{eqn:omega-GW-eccentric})--(\ref{eqn:u-eccentric}), and the column ``$M\Omega$'' in Table~\ref{tbl:NRparams} reports the \emph{mean-motion} $Mn$, cf. Eq.~(\ref{eqn:meanmotion}), at the reference epoch.  Furthermore, for these eccentric simulations, the eccentricity is reported at the same frequency $Mn=0.0272712$ for all simulations, corresponding to a (2,2) \GW frequency of $f_{\rm GW}=23.8$~Hz.  This decouples the value of the eccentricity from the individual starting frequency of each eccentric simulation (recall that orbital eccentricity decays during the inspiral~\cite{peters:1964}).

The final column of Tables~\ref{tbl:NRparams}
and~\ref{tbl:NRparams_additional} indicates the numerical truncation
error of the simulations computed as follows. For each \NR simulation, we take the two waveforms with the
highest resolutions at an inclination of $\iota=\pi/3$ and compute the noise-weighted inner product between
them.
More precisely, we follow the approach of Refs.~\cite{Damour:1997ub,Babak:2016tgq} by considering the notion of the min-max overlap that
gives the lowest overlap when considering all sky positions and
polarizations.

Specifically, given a waveform of one resolution, $h_1$, evaluated at a fixed set of parameters, we
choose the polarization angle and sky location of the other resolution $h_2$,
such that the overlap given by
\begin{equation}
\mathcal{O}(h_1,h_2):= \frac{(h_1\,|\,h_2)}{\sqrt{(h_1\,|\,h_1)(h_2\,|\,h_2)}}
\end{equation}
between the waveforms of the two numerical resolutions is maximized,
where the inner product $(.\,|\, .)$ is defined in
Eq.~(\ref{eq:inner-product}). In addition, we also maximize the
overlap over a time- and phase-shift between the two waveforms. We
then minimize the overlap over the sky location and polarization of
$h_{1}$. By construction, the overlap will always be equal to or above
the min-max, regardless of the source parameters, thus making it a
suitable conservative measure.
An overlap of $\mathcal{O} = 1$ indicates perfect agreement between two
waveforms.  The deviation of the overlap from one, $1 - \mathcal{O}$,
is referred to as \emph{mismatch}, and is a useful measure to
approximately quantify the accuracy of waveforms.

This quantity is averaged over several azimuthal angles and is
reported in the last column as $1 - \mathcal{O}_{\rm res}$ in
Tables~\ref{tbl:NRparams} and~\ref{tbl:NRparams_additional}.

In the Fisher-matrix approximation for the single-detector
case, two waveforms are considered indistinguishable if their
mismatch satisfies
$1 - \mathcal{O} \lesssim 1/(2\rho^2)$~\cite{Flanagan:1997kp,lindblom:2008cm,McWilliams:2010eq}.
For $\rho=25$, this implies that
errors in the numerical waveforms will be irrelevant if they lead to
mismatches $\lesssim 8\times 10^{-4}$. For the numerical truncation
error (as considered in the column $1 - \mathcal{O}_{\rm res}$ in
Tables~\ref{tbl:NRparams} and~\ref{tbl:NRparams_additional}), we
find that this is the case for the numerical simulations considered
here. For the SXS waveforms, a detailed analysis of other sources
of errors in the numerical simulations finds that other sources of
error dominate over numerical truncation error, most notably
ambiguities in gravitational-wave extraction, however, the combined
error still leads to mismatches $\lesssim 3\times 10^{-4}$. Therefore, we conclude that the SXS simulations are sufficiently accurate for the present study, a conclusion also confirmed in Sec.~\ref{sub:effect_of_numerical_errors} below.

\subsection{Bayesian parameter estimation}
\label{sub:bayesian_parameter_estimation}

The \emph{posterior probability density function} (PDF) of a set of parameters $\boldsymbol{\theta}$
which describe the physical properties and
orientation of the binary system can be expressed with Bayes' theorem~\cite{Bayes:1793, Jaynes:2003},
\begin{equation}
\label{eqn:posterior_prob}
p(\boldsymbol{\theta}\,|\,d, H) = \frac{p(\boldsymbol{\theta}\,|\,H)\,\Lambda(d\,|\,\boldsymbol{\theta}, H)}{\int
p(\boldsymbol{\theta}\,|\,H)\,\Lambda(d\,|\,\boldsymbol{\theta}, H)\;d\boldsymbol{\theta}},
\end{equation}
where $p(\boldsymbol{\theta}\,|\,H)$ is the \emph{prior probability density} for
$\boldsymbol{\theta}$ given a model $H$ and $\Lambda(d\,|\,\boldsymbol{\theta}, H)$ is the
\emph{likelihood} function. In the case of \GW data, the data $d$ is
described by the signal $h(\boldsymbol{\theta}_0)$ with given parameters $\boldsymbol{\theta}_0$
and the instrument noise $n$. The likelihood function can then be
written as~\cite{Cutler:1994ys,Veitch:2014wba}
\begin{equation}
\label{eqn:likelihood_ratio}
\Lambda(d|\boldsymbol{\theta}) \propto \exp\left( -\frac{1}{2}\left(h(\boldsymbol{\theta})-d \left|\right.
h(\boldsymbol{\theta})-d\right) \right),
\end{equation}
where the notation $(a\,|\,b)$ indicates the noise-weighted inner
product, cf. Eq.~(\ref{eq:inner-product}).

In order to measure the recovered distribution of the binary system properties,
we inject the waveform with the given set of parameters into the data,
and use two independent stochastic samplers, based on
parallel-tempered Markov-Chain Monte Carlo (MCMC) and on nested
sampling algorithms. Our set-up is consistent
with~\cite{PhysRevLett.116.241102}, and the engine implementations are
available in the \texttt{LALInference} package \cite{Veitch:2014wba}
of the \ac{LAL} software suite~\cite{lal}.

The samplers are specially designed for \GW data
analysis, and as well as generating posterior samples for the waveform
parameters, they are also capable of marginalizing over uncertainties in
the posteriors propagated from the uncertainties in the model used to
calibrate \GW strain data~\cite{GW150914-DETCHAR}. The marginalization
assumes that errors in the phase and amplitude of the data can be fit
with a spline model consisting of $\sim5$ points placed at intervals in
the frequency domain~\cite{SplineCalMarg,PhysRevLett.116.241102}.

To represent the full joint distribution of the parameters would be
unfeasible, so instead, we present posteriors marginalized in all but
one or two dimensions: the width of these posteriors (often encoded in a
confidence interval) encodes the statistical uncertainty in the measurement.
However, it is important to note that many of the parameters have correlated
probability densities (e.g., distance/inclination, component masses).

For the most part of this study we do not include noise in the simulated data in order
to focus on comparing systematic against statistical errors in an idealized setting.
If a waveform model were a perfect match for an \NR signal, then the noise-free analysis
should yield a posterior \PDF peaked at the true parameter values up to biases induced
by the priors.

Including detector noise as in the analysis presented in Sec.~\ref{sub:effect_of_detector_noise}
will smear out and shift the posteriors and allow us to get a sense of realistic statistical
uncertainties. The presence of noise will also reduce the impact of systematic biases inherent
in waveform models and therefore the noise-free analysis should be conservative.


\section{Results}
\label{sec:results}

\TheEvent{} has been shown to be consistent with a range of source parameters
\cite{TheLIGOScientific:2016pea,PhysRevLett.116.241102,Abbott:2016izl}, and below we shall
show how reliably the methods and waveform models described here can extract the
properties of signals that are consistent with the parameter estimates of \TheEvent{}.
We first analyze non-precessing signals and confirm that all models
give reliable results, as expected from their tuning to non-precessing \ac{NR}
simulations \cite{Khan:2015jqa,Husa:2015iqa,taracchini:2012ig,Kumar:2015tha},
and the comparatively small amplitude of higher harmonics for almost equal mass ratios,
spins and orientations.
In Sec. \ref{sub:precession}, we turn to precessing signals. Generally, we find no
significant parameter biases except for particular choices of polarization angle
and source inclination. We discuss this effect in more detail, and demonstrate that
systematic biases would be significant for only a small fraction of possible source
orientations, and we can confidently conclude that the analysis of \TheEvent{}
did not suffer from these biases.

\subsection{Non-precessing binaries}
\label{sub:aligned_spin}

We first study the parameter recovery for \ac{NR} waveforms with \BH spins
aligned with the orbital angular momentum direction $\hat L$ of the
binary. The physical effect of aligned (anti-aligned) spins is to increase
(decrease) the number of orbits accumulated from a reference frequency
to merger relative to a non-spinning binary.
We inject a number of aligned-spin \NR waveforms into zero noise and use non-precessing
EOBNR~\cite{Taracchini:2013rva,Purrer:2015tud} and non-precessing IMRPhenom
waveforms~\cite{Khan:2015jqa} to estimate the source parameters.

\begin{figure*}[h!tb]
  \centering
\includegraphics[scale=0.67,trim=0 8 0 10,clip=true]
{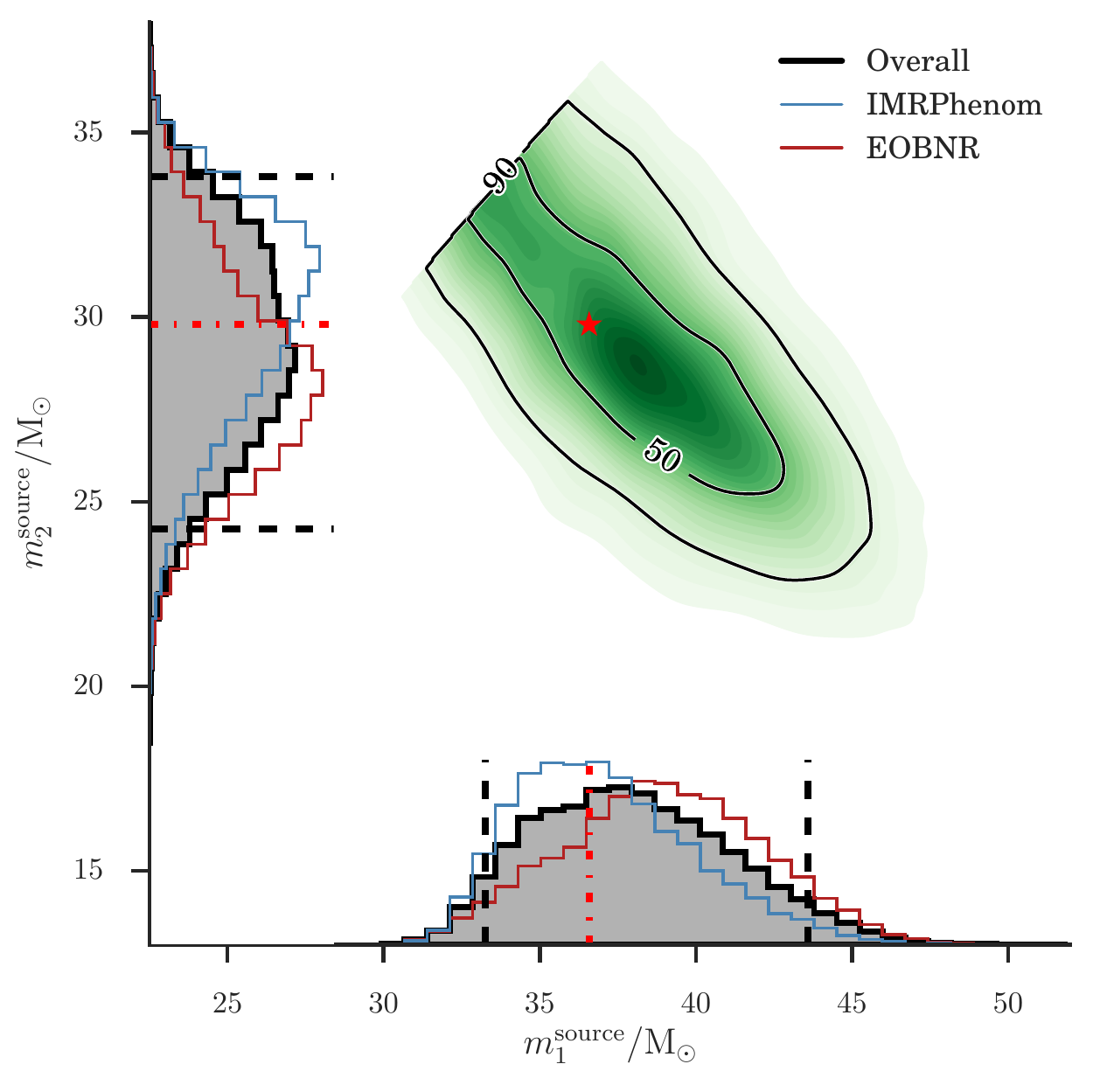}
$\;\;\;\;\;\;\,$
\includegraphics[scale=0.67,trim=0 8 0 10,clip=true]
{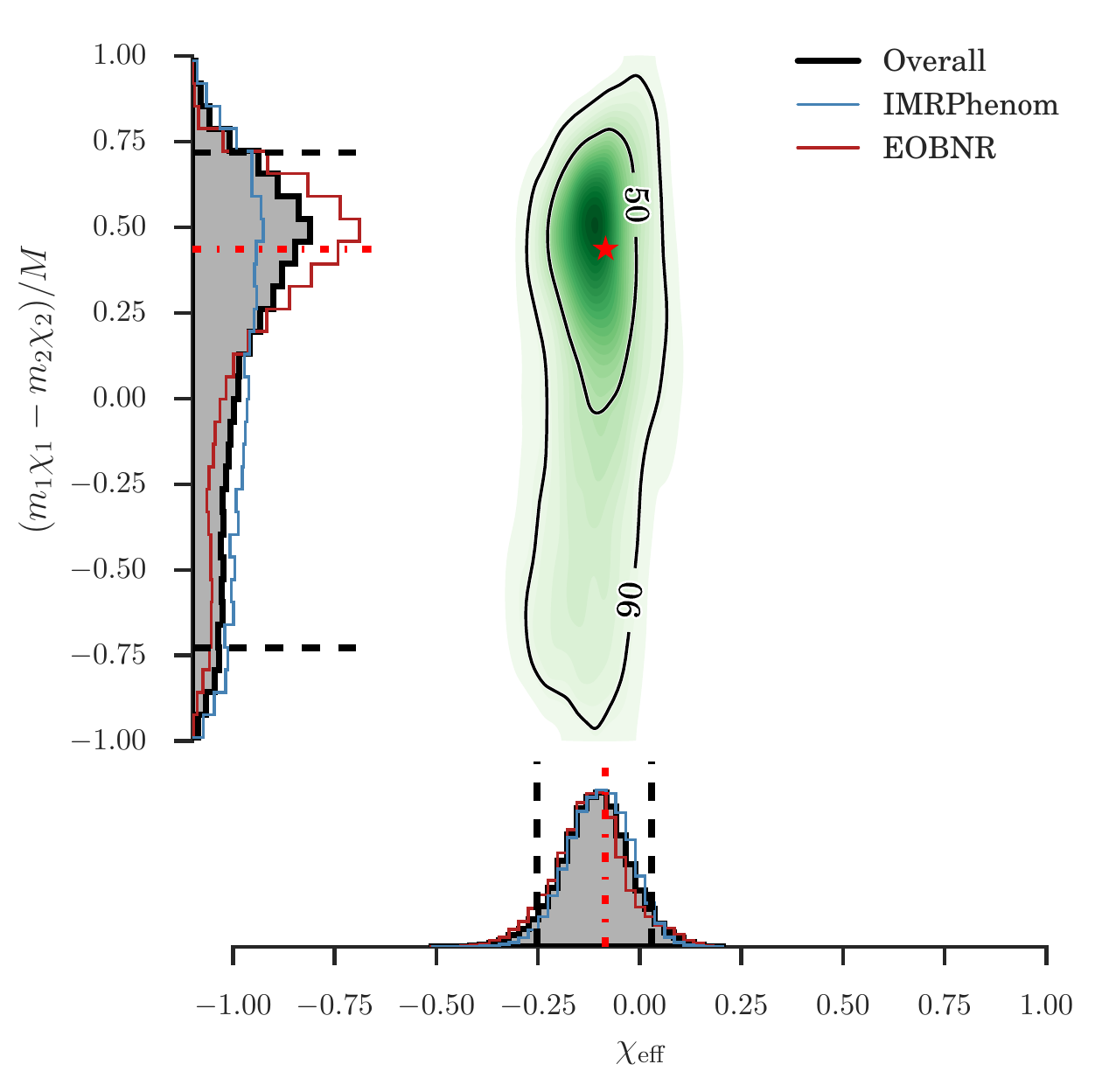}
\caption{
  Comparison of source frame component masses and aligned spin combinations for an aligned
  \NR mock signal (\texttt{SXS:BBH:0307}) with masses and spins consistent with~\TheEvent{}.
  The signal is injected into zero noise using the fiducial inclination, $\iota =
  163\degree$, and polarization angle $\psi = 82\degree$. The non-precessing IMRPhenom and
  EOBNR models are used for recovery. The left panel shows credible regions for recovery of
  the component masses, whereas the right panel shows spin recovery.
  As in~\cite{PhysRevLett.116.241102} we combine the posterior samples of both models
  with equal weight, in effect marginalizing over our choice of waveform model. The
  resulting posterior is shown in the two-dimensional plot as the contours of the $50\%$
  and $90\%$ credible regions plotted over a color-coded \PDF.
  Dashed lines in the one-dimensional plots show $90\%$ credible intervals of the
  individual and combined posteriors.
  The injected parameter values are shown as red dot-dashed lines and a red asterisk.
  Both models recover the correct masses and effective spin $\chi_{\rm eff}$ within
  the $90\%$ credible regions, while the anti-symmetric spin combination is
  not measured well; the peak in the EOBNR \PDF around the correct value is a
  spurious effect (see text).
}
  \label{fig:plots_SXS_Ossokine_0233_HM}
\end{figure*}

Fig.~\ref{fig:plots_SXS_Ossokine_0233_HM} shows marginalized posterior \acp{PDF}
for the spin and mass parameters for both waveform models
for the \NR signal \texttt{SXS:BBH:0307} with intrinsic parameters as listed in Table~\ref{tbl:NRparams} and fiducial parameters listed in Table~\ref{tab:fiducial}.
We find that the true parameter values lie well within the 90\% credible intervals
for either model for these fiducial values.
Below we quote the medians, $90\%$ credible intervals and an estimate for the $90\%$ range of
the systematic error determined from the variance between the two waveform models.
As is expected from earlier theoretical studies~\cite{aasi:2014tra,Kumar:2016dhh},
and consistent with previous LVC studies of \TheEvent{}, we find excellent agreement in the
chirp mass, given by
\begin{equation}
\label{eq:m_chirp}
\mathcal{M} = \frac{(m_1 m_2)^{3/5}}{M^{1/5}},
\end{equation}
which is the coefficient of the leading-order term in the \ac{PN} phase evolution,
while the mass ratio $q$ is broadly consistent with the injected value.
For the heavy \BBH systems considered here the total mass of the binary is similarly well constrained as the chirp mass~\cite{Veitch:2015ela,Graff:2015bba} due to the dependence of the ringdown on the total mass~\cite{Berti:2005ys}.
The difference in the location of the peaks in the \PDF for the component masses
(left panel of Fig.~\ref{fig:plots_SXS_Ossokine_0233_HM}) is
due to small differences between the two waveform models.
For the source frame masses\footnote{We measure redshifted masses $m$, which are related to source-frame masses using the relation $m = (1+z) m^\mathrm{source}$~\cite{Krolak:1987ofj}, where $z$ is the cosmological redshift.} we find
$m_1^\mathrm{source} = 37.8_{- 4.5 \pm 0.5}^{+ 5.8 \pm 1.6}$~$M_\odot$ and
$m_2^\mathrm{source} = 29.2_{- 5.0 \pm 0.9}^{+ 4.6 \pm 0.7}$~$M_\odot$,
and see that the systematic errors are about a factor 5 smaller than the statistical errors.
The effective spin recovery is consistent between the two models,
$\chi_\mathrm{eff} = -0.11_{- 0.15 \pm 0.04}^{+ 0.14 \pm 0.01}$.
However, an anti-symmetric combination of the two spins is not well constrained,
indicating the difficulty in measuring the difference between the two spins~\cite{purrer:2015nkh},
$(m_1\chi_{1L} - m_2\chi_{2L})/M = 0.29_{- 1.02 \pm 0.16}^{+ 0.43 \pm 0.05}$.
We note that EOBNR leads to a markedly more pronounced peak of the anti-symmetric $(m_1\chi_{1L} - m_2\chi_{2L})/M$ posterior. The EOBNR model incorporates both spins $\chi_{1L}, \chi_{2L}$, whereas IMRPhenom primarily utilizes the effective spin $\chi_\mathrm{eff}$.
Therefore, the improved recovery of the anti-symmetric spin combination with EOBNR points to some extra power afforded by the more complete model.
However, this improved recovery is not a generic feature found in other configurations and seems to be a spurious effect. This is supported by Ref.~\cite{Kumar:2016dhh} who find a mismatch of $\sim 3\%$ for non-precessing EOBNR for configurations with highly anti-symmetric spins at equal-mass. In addition, our analysis uses idealized assumptions of injections into zero-noise, while PE analyses in non-Gaussian detector noise and marginalization over calibration errors would wash out such fine features and their extraction would require SNRs much higher than 25.

In addition, we performed injection and parameter-recovery for a large number of non-precessing-binary signals listed in Table~\ref{tbl:NRparams_additional} with
similar results, as summarized in Table~\ref{tab:aligned_summary} in Appendix~\ref{sec:parameter_estimation_results_for_additional_nr_configurations}.
In summary, we recover parameters that are statistically consistent between the EOBNR and
IMRPhenom models, and with those describing the mock \ac{NR} source. These results
confirm previous studies~\cite{Khan:2015jqa,Kumar:2016dhh}.

\subsection{Precessing binaries}
\label{sub:precession}

\subsubsection{Fiducial inclination and polarization}
\label{ssub:precession_weak_modulation}

\TheEvent{} is consistent with a wide range of \BH spin configurations, including the possibility that one or both \BH spins are misaligned with the orbital angular momentum. Such misalignments
give rise to precession of the \BH spins and the orbital plane of the binary, leading to modulations in the gravitational waveform~\cite{apostolatos:1994, kidder:1995zr}. We now explore the parameter recovery of such precessing sources.

First, we analyze precessing \ac{NR} signals injected with the fiducial parameters listed in Table~\ref{tab:fiducial}.
While the inclination $\iota$ is time-dependent for precessing binaries, the orientation of the total orbital angular momentum $\hat{\mathbf{J}}$
remains almost constant.\footnote{The exception to this are binaries that undergo transitional precession~\cite{apostolatos:1994}.} It is therefore often more meaningful to consider the angle $\theta_\mathrm{JN}$ between $\hat{\mathbf{J}}$ and the line-of-sight $\hat{\mathbf{N}}$ instead. We note that for the precessing binaries discussed in this study the opening angle of the precession cone of $\hat{\mathbf{L}}$ around $\hat{\mathbf{J}}$ is only a few degrees at $30$~Hz and thus $\theta_\mathrm{JN}$ and $\iota$ are close.

As in the original analysis of the properties of \TheEvent{} in Ref.~\cite{PhysRevLett.116.241102}, we use the non-precessing EOBNR and
precessing IMRPhenom waveform models in this study. Analyses with the precessing EOBNR model are currently not computationally feasible
to perform detailed investigations. A comparison between the two precessing models in the estimation of the properties of \TheEvent{} and against two \ac{NR} injections is discussed in
Ref.~\cite{Abbott:2016izl}. It found that the two precessing models showed good agreement in the recovery of both injections.

\begin{figure}[htbp]
  \centering
  \includegraphics[scale=0.67,trim=0 8 0 10,clip=true]{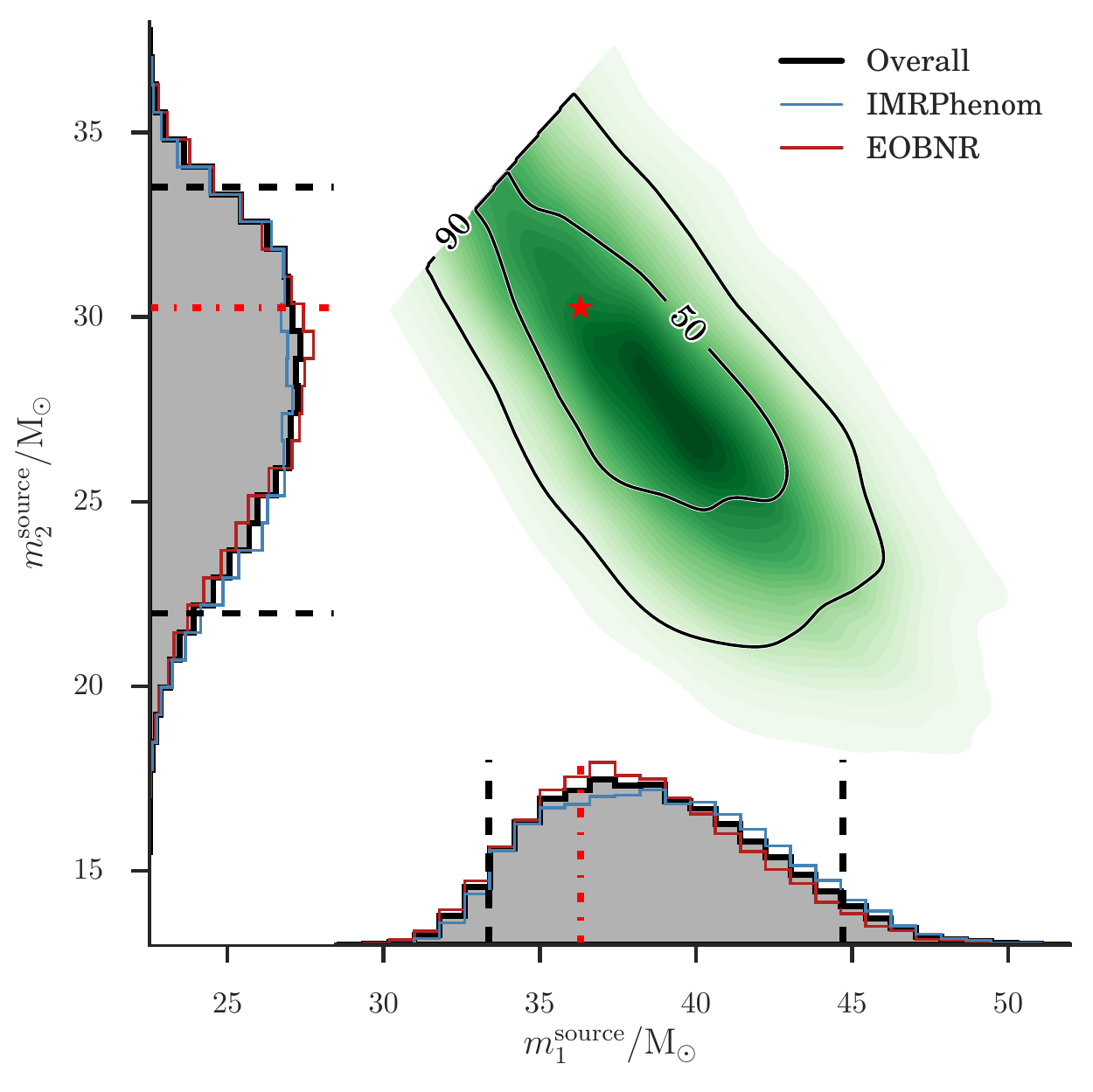}
  \caption{
  Comparison of component masses for a precessing NR mock signal (\texttt{CFUIB0029})
  with masses and spins consistent with \TheEvent{}. The mock signal is injected in
  zero noise using the fiducial inclination, $\iota = 163\degree$,
  and polarization angle $\psi = 82\degree$. The precessing IMRPhenom and non-precessing
  EOBNR models are used for recovery.
  }
  \label{fig:plots_mass_NR_SO_prec}
\end{figure}

\begin{figure*}[htbp]
  \centering
  \includegraphics[scale=0.6,trim=0 8 0 16,clip=true]{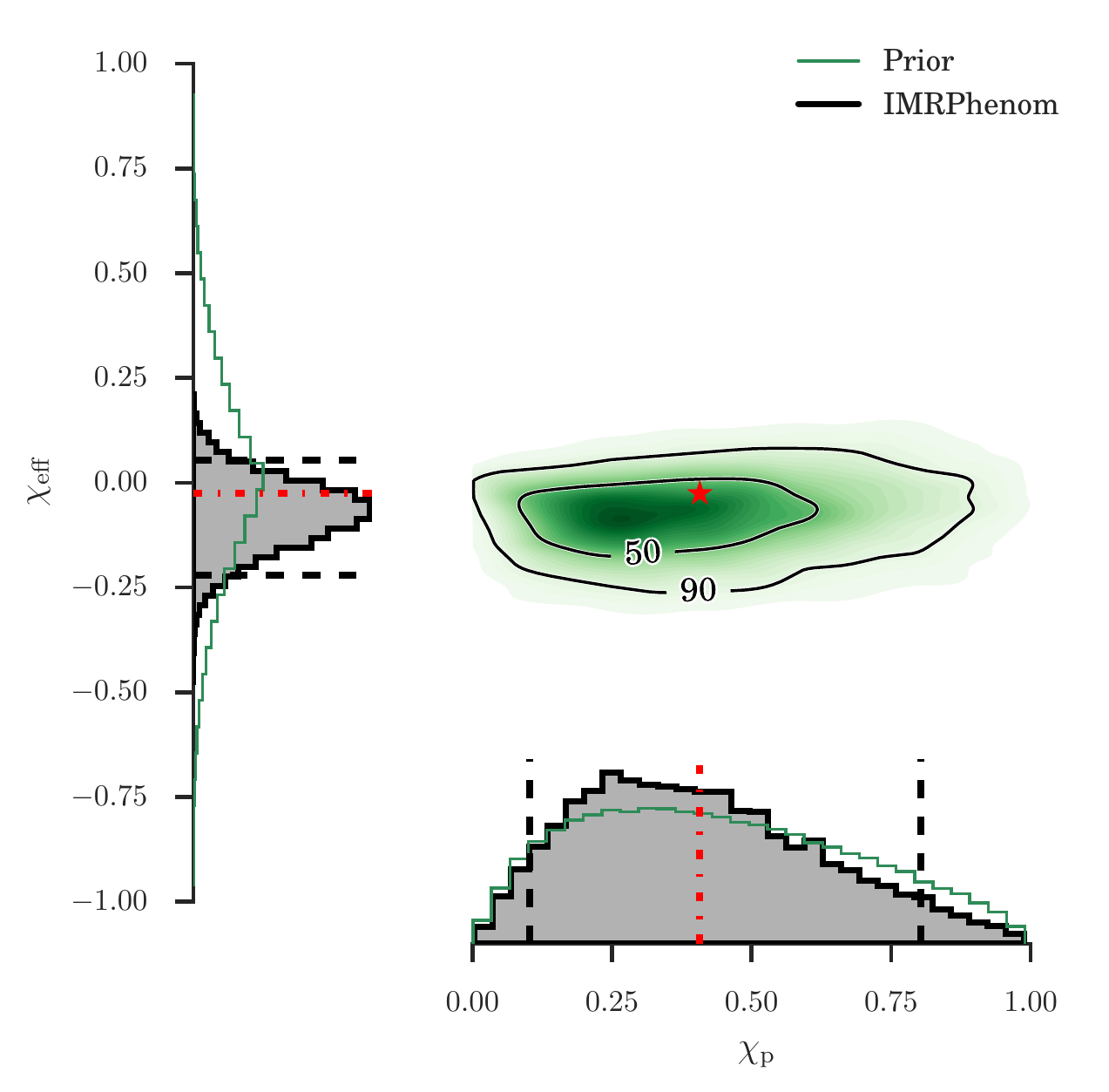}
\hspace*{3em}
\includegraphics[scale=0.6,trim=0 4 0 5,clip=true]{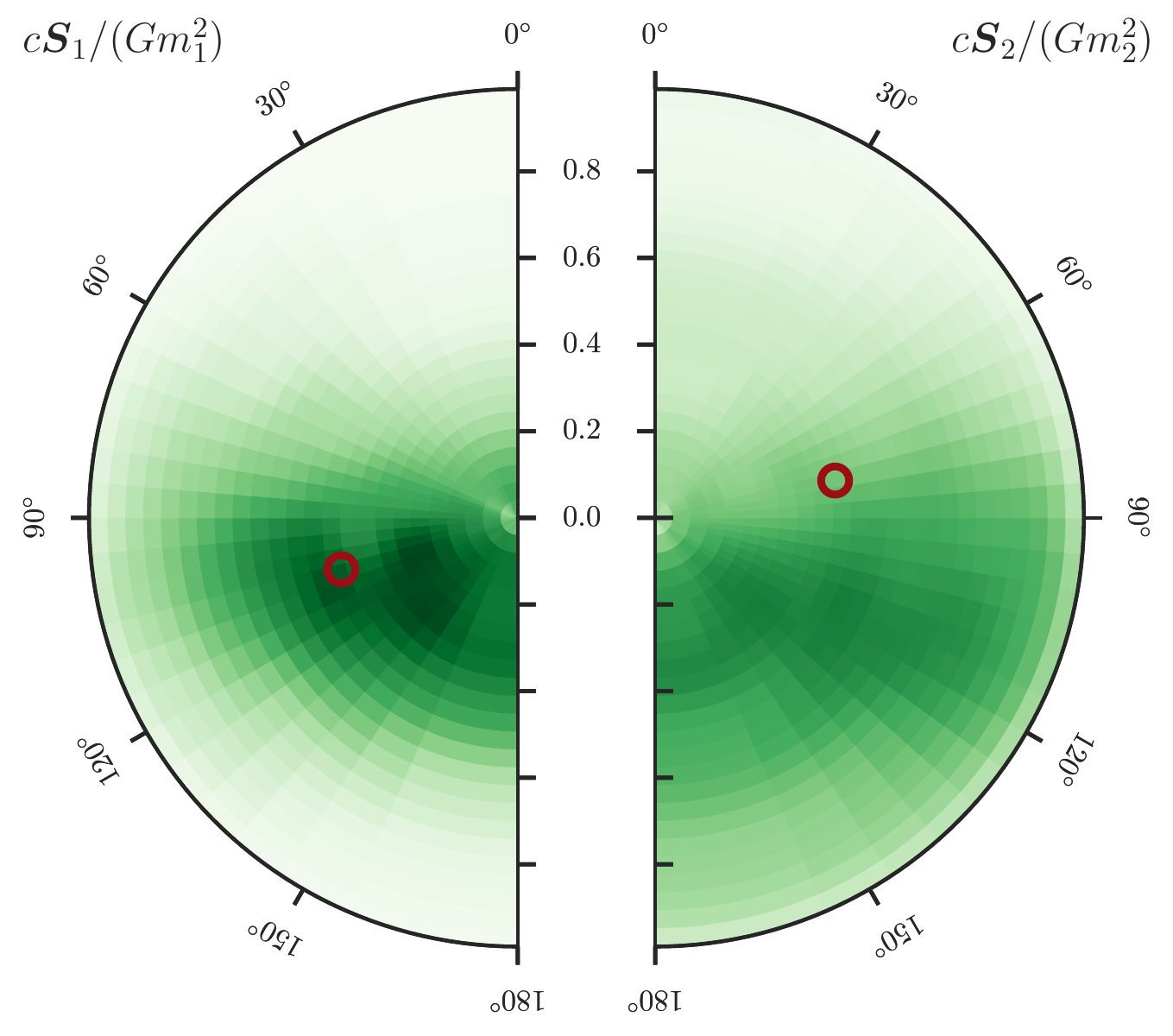}
  \caption{
  Comparison of spins for the precessing NR mock signal (\texttt{CFUIB0029}) shown
  in Fig.~\ref{fig:plots_mass_NR_SO_prec}.
  Left: PDFs for the $\chi_\mathrm{p}$ and $\chi_\mathrm{eff}$ spin parameters. The one-dimensional plots show probability contours of the prior (green) and marginalized \PDF (black).
  The dashed vertical lines mark $90\%$ credible intervals. The two-dimensional plot shows the contours of the $50\%$ and $90\%$ credible regions plotted over a color-coded \PDF. The injected parameter values are shown as red dot-dashed lines and a red asterisk.
  Right: PDFs for the dimensionless component spins $c\boldsymbol{S}_{1}/(Gm_1^2)$ and $c\boldsymbol{S}_{2}/(Gm_2^2)$ relative to the normal to the orbital plane $\boldsymbol{L}$, marginalized over uncertainties in the azimuthal angles.
   }
  \label{fig:plots_spins_NR_SO_prec}
\end{figure*}

For this first study we choose \texttt{CFUIB0029} (see Table~\ref{tbl:NRparams}),
a simulation where the BH spins point predominantly in the orbital plane, and with a
reasonably large value of $\chi_p \approx 0.4$. We inject this waveform at fiducial
parameters into zero noise. Figs.~\ref{fig:plots_mass_NR_SO_prec} and~\ref{fig:plots_spins_NR_SO_prec} summarize the parameter recovery for this injection.
We find that the true parameter values of the \ac{NR} signal (red asterisks) lie within the
$50\%$ credible regions for component masses and effective spins indicating unbiased parameter recovery for this injection with either waveform model.
For the source frame masses we find
$m_1^\mathrm{source} = 38.3_{- 4.9 \pm 0.3}^{+ 6.4 \pm 0.7}$~$M_\odot$ and
$m_2^\mathrm{source} = 28.2_{- 6.2 \pm 0.4}^{+ 5.3 \pm 0.3}$~$ M_\odot$,
with systematic errors an order of magnitude smaller than statistical errors.
For the effective aligned spin we have
$\chi_\mathrm{eff} = -0.08_{- 0.19 \pm 0.06}^{+ 0.15 \pm 0.02}$.
Here systematic errors are a factor four smaller than statistical errors.
The absolute bias between the true parameter values and the overall medians in the source frame masses is $\approx 2 M_\odot$ and $\approx 0.05$ in $\chi_\mathrm{eff}$.
The spin directions as shown in the right panel of Fig.~\ref{fig:plots_spins_NR_SO_prec} are
not constrained. No information on the effective precession spin
$\chi_p$ is recovered, despite the signal having appreciable
$\chi_p$. Instead, we effectively recover the prior on $\chi_p$ as can be seen
in the left panel of Fig.~\ref{fig:plots_spins_NR_SO_prec}.
This may be attributed to the following reasons: Firstly, the fiducial inclination
only gives rise to weak precession-induced modulations in the signal, and secondly
the shortness of the signal only allows for at most one modulation cycle in the aLIGO
sensitivity window.
Hence we find that for the fiducial parameters, parameter recovery is
\emph{not biased} in the sense that the injected values are always well inside their
posterior confidence regions.

Parameter estimates were obtained for several additional NR signals in the vicinity of \TheEvent{} with the precessing IMRPhenom model for fiducial and also edge-on inclinations of the source. The results are summarized in Table~\ref{tab:precessing_summary} in Appendix~\ref{sec:parameter_estimation_results_for_additional_nr_configurations}. These results agree with
the findings in this section that parameter recovery is not biased for the fiducial parameters.
On the other hand, if the source is viewed at nearly edge-on, inclination biases can arise and we will discuss these next in Sec.~\ref{sec:effects_of_pol}.
We note that for some configurations we find small disagreements in the shapes of the \acp{PDF}, similar to those found for
the non-precessing injection in Fig.~\ref{fig:plots_SXS_Ossokine_0233_HM}. However, these differences do not noticeably affect
the credible intervals, and we find no clear relationship between the level of disagreement and the location in parameter space.

\subsubsection{Varying inclination and polarization}
\label{sec:effects_of_pol}

\begin{figure*}[htbp]
    \includegraphics[width=.44\textwidth,clip=true,trim=0 16 0 12]{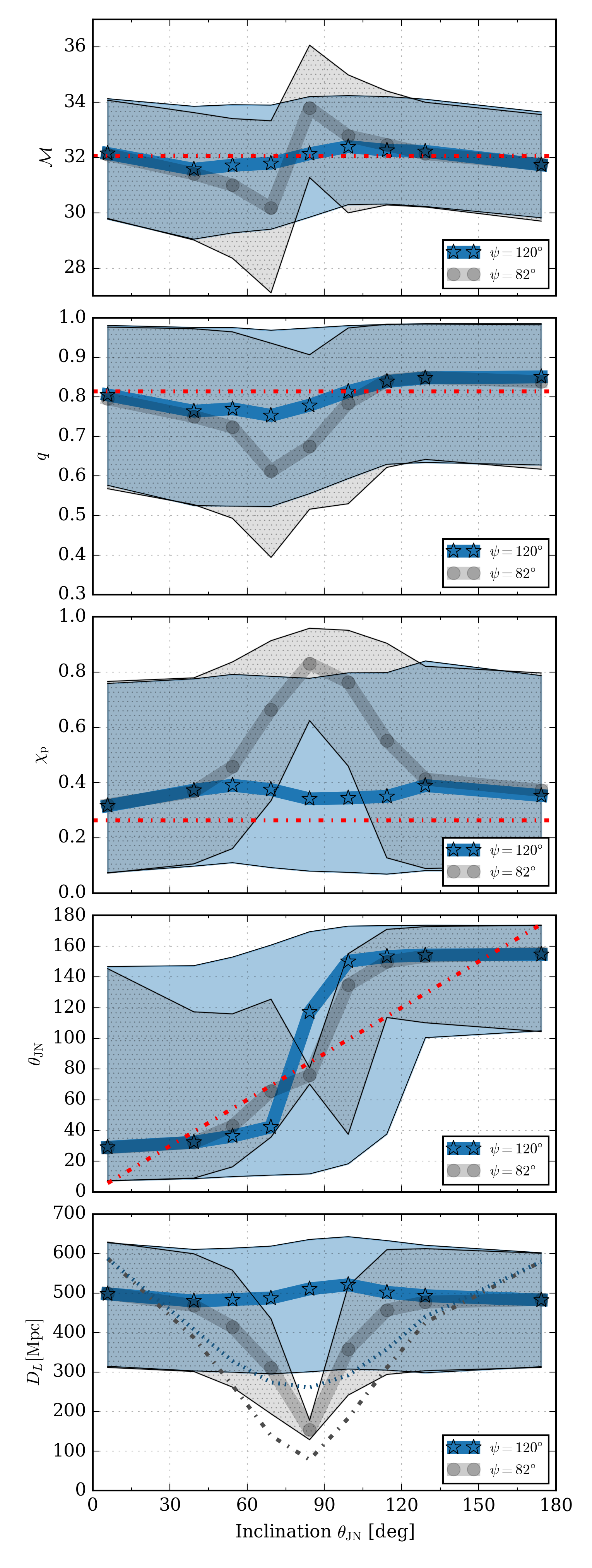}
    \includegraphics[width=.4425\textwidth,clip=true,trim=0 16 0 12]{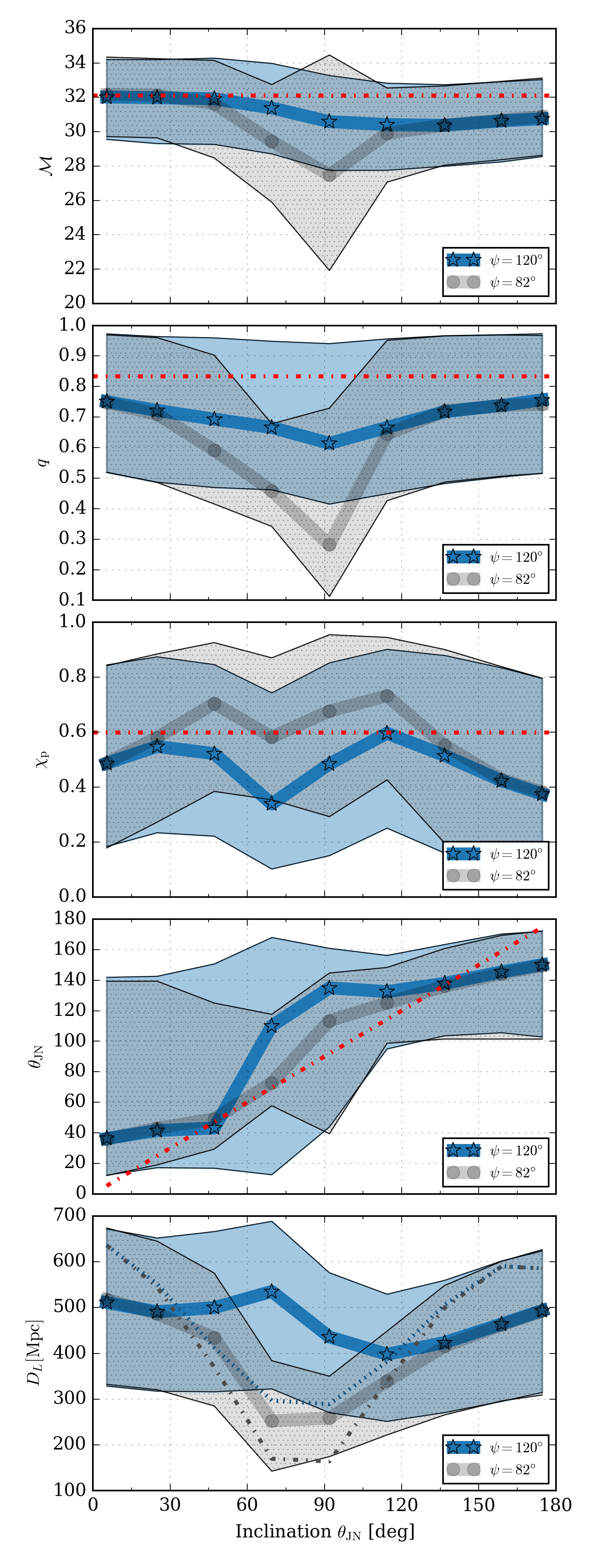}
  \caption{
  Inclination dependence of parameter recovery. Two NR waveforms primarily differing in $\chi_p$ (\texttt{SXS:BBH:0308} in left column; \texttt{CFUIB0020} in right column) are injected with different $\theta_\mathrm{JN}$ as given on the ordinate axes. Shown on the abscissa axes
are $90 \%$ credible intervals (blue / gray bands) and medians (asterisks / circles) for these precessing NR signals recovered with the precessing IMRPhenom model.
Injected parameter values are shown as red dash-dotted lines, except for the bottom two panels where the injected values depend on $\psi$ and are shown in blue (dotted) and gray (dash-dotted).
Shown from top to bottom are
chirp mass $\mathcal{M}$, mass-ratio $q$, effective precession spin $\chi_\mathrm{p}$, the angle $\theta_\mathrm{JN}$ and luminosity distance $D_L$.
The analysis is repeated for two choices of detector polarization angle $\psi$, with the one shown in grey representing a detector orientation approximately canceling $h_+$.
  }
  \label{fig:phenp_posteriors}
\end{figure*}

In Sec.~\ref{ssub:precession_weak_modulation} we found that the IMRPhenom and EOBNR models recover the injected parameters with comparable accuracy as expected from the results of \TheEvent{}, without significant bias. However, precession-induced signal modulations become stronger for sources viewed at an angle of $\theta_\mathrm{JN} \sim 90\degree$ (edge-on)(see e.g.~\cite{apostolatos:1994, arun:2008kb, schmidt:2012rh}).
For such orientations two qualitatively new features arise in the waveform:
the circular orbital motion becomes approximately linear when observed edge-on, thus preferring the observation of the plus polarization, while the precession of the orbital plane dominates the other polarization. Signals from such sources have a richer waveform structure and are more challenging to capture by the models discussed here.
When injecting and recovering precessing waveforms edge-on, we find:
(1) \PE may yield biased results with the level of bias depending on
both the source inclination \emph{and} signal polarization;
(2) the bias is most likely caused by discrepancies between precessing IMRPhenom
and the fully general-relativistic NR signals, but
(3) these biases only manifest themselves for certain source orientations and polarizations,
and as such are likely to constitute only a small fraction
of observations.

The inclination $\iota$ of the source relative to the detector strongly affects the morphology of the detected signal,
In addition, the signal recorded at the detector also depends on the polarization angle $\psi$ as well as the position in the sky $(\alpha, \delta)$ (cf. Eq.~(\ref{eq:detector_response})).
For the current two-detector network, which is principally sensitive to only one \GW polarization for any given sky location,
this suggests that $\psi$, $\alpha$ and $\delta$ may be partially degenerate, as supported by Refs.~\cite{Finn:1992xs,raymond:2014uha}.
Therefore, we expect that varying $\psi$, while fixing the sky position, will lead to an effective exploration of the extrinsic parameter space.
In this context, we assess how well the precessing IMRPhenom waveform model approximates \GW{} signals when varying amounts of $h_+$ and $h_{\times}$ polarization are present at different inclinations.

We focus our investigation on four NR simulations: \texttt{CFUIB0029, CFUIB0012} and \texttt{CFUIB0020}, and \texttt{SXS:BBH:0308} as listed in Table \ref{tbl:NRparams}.
We find the results to be qualitatively consistent between all
four cases, and in what follows we focus on \texttt{CFUIB0020} and \texttt{SXS:BBH:0308}, as examples of waveforms from two independent
\ac{NR} codes.
\ac{NR} injections were again performed into zero noise with the fiducial parameter given in Table~\ref{tab:fiducial} but with varying inclination and polarization angles.
An overview of our results is given in Fig.~\ref{fig:phenp_posteriors}.
We find that:
\begin{itemize}
	\item[$\circ$] Results are qualitatively similar between simulations (\texttt{CFUIB0020}
  and \texttt{SXS:BBH:0308}) for two different choices of physical \BBH parameters.
	\item[$\circ$] Parameter estimates with IMRPhenom are most accurate for
  signals with inclinations near $0\degree$ (``face-on'') or $180\degree$ (``face-off'').
	\item[$\circ$] Results depend on the polarization angle when signals
 have an inclination near $90\degree$.
	\item[$\circ$] For inclinations and polarization angles in a region near
  $90\degree$, recovered parameter values (e.g., for mass ratio) deviate most
  strongly from injected parameters. In rare cases, the injected parameters
  lie outside the $90\%$ credible region.
\end{itemize}

\begin{figure*}[htbp]
  \centering
  \includegraphics[width=.96\textwidth,trim=0 10 0 4,clip=true]{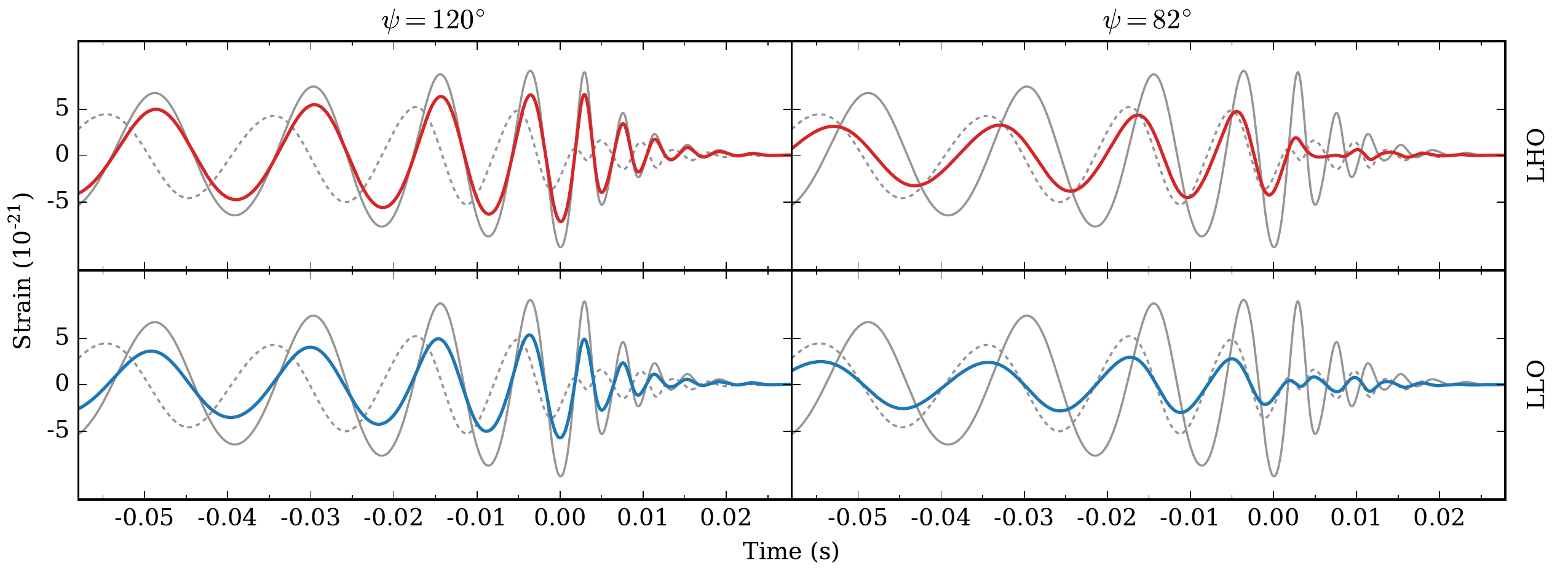}
  \caption{
    Comparison of detector responses using \ac{NR} waveform \texttt{CFUIB0020}
at an angle $\theta_\mathrm{JN}=92\degree$. Incoming $h_{+}$ and $h_{\times}$
polarizations are shown by grey solid and dashed curves respectively. The left column shows
the theoretical detector response (Eq.~(\ref{eq:detector_response})) in red for \ac{LHO} on
the top row and in blue for \ac{LLO} on the bottom row, for a polarization
angle of $\psi=120\degree$. The right column shows the responses for a
polarization angle of $\psi=82\degree$. In both cases the distance
to the source is 40 Mpc. The signal at \ac{LHO} has been inverted to account for
the relative detector orientations, and the signals are aligned in time; cf.\
Fig.~1 in \cite{Abbott:2016blz}. We find significant differences in the detector
responses between the two polarization angles (see text for details).
  }
  \label{fig:td_response}
\end{figure*}

While Fig.~\ref{fig:phenp_posteriors} demonstrates qualitatively similar results for the
parameter recovery of the \texttt{SpEC} and the BAM simulations, we note two differences
for near edge-on inclination at $\psi=82\degree$: For \texttt{SXS:BBH:0308} (left column)
the recovered distance is overestimated, correlated with a significantly overestimated
$\chi_p$, while the other parameters show no large biases. For \texttt{CFUIB0020}
(right column), on the other hand, the mass ratio $q$ shows a significant bias.
We find that both $\chi_p$ and $q$ are correlated with extrinsic parameters. Depending on details of the \NR signal, either one can be heavily biased.
We further point out that for this orientation and polarization:
(a) The distance prior (strongly) and the likelihood (less strongly) favor a source-distance larger than the injected distance.
(b) The posterior samples show systematic biases: Those samples near the actual injected distance correlate with less biased values of $\chi_p$ and $q$ than posterior samples that have a larger distance.
(c) The recovered sky position $(\alpha, \delta)$ is strongly biased.
(d) The recovered SNR is only 23 (for the injection at SNR of 25).

Conversely, we see no biases when the source polarization matches the
\ac{DPF}~\cite{klimenko:2005wa,Sutton:2009gi} of the network, for $\psi \sim 120\degree$,
(i.e., when the network has maximum response to the source-frame $h_+$ and minimum
response to the source-frame $h_\times$).
However, we see clear biases when the source polarization
is near $45 \degree$ to the \ac{DPF} (i.e., when the network has maximum response to the
source-frame $h_\times$ and minimum response to the source-frame $h_+$) as described above.

We can see that this is indeed the case by considering the time-domain
waveforms for cases with and without observed parameter biases. Fig.~\ref{fig:td_response} shows
the detector response $h_\mathrm{resp}$ and the incident \GW polarizations, $h_+$
and $h_\times$, for LIGO-Hanford and LIGO-Livingston for \texttt{CFUIB0020}
viewed edge-on with $\theta_\mathrm{JN}=92\degree$. The components $h_+$ and
$h_\times$ are the same on the left and right panels, but the proportion of each
polarization that contributes to the signal $h_\mathrm{resp}$ differs.
The left panels show the complete signal for a polarization angle $\psi=120\degree$, and the
right panels show the complete signal for a polarization angle $\psi=82\degree$. We see that for $\psi = 120\degree$ the observed signal is
dominated by $h_+$, whereas in the right panels the observed signal is dominated by $h_\times$.

In particular, we see that for a given polarization angle and source inclination, the detector response may correspond to the partially constructive or destructive interference of $h_+$ and $h_\times$, which amplifies or diminishes the observed signal. Such cases are especially challenging and require waveform models that describe $h_{+,\times}$ very accurately. However, in the construction of the precessing IMRPhenom only the aligned-spin $(2,\pm 2)$-modes are used.
By construction, the neglect of higher-order aligned-spin modes results in approximate precessing modes. This approximation becomes
more inaccurate for systems close to edge-on, as contributions from higher-order modes to the observed signal become more important. In addition, for a suitable polarization angle $\psi$, the $(2,\pm 2)$-contributions to the signal
may vanish completely and any observed strain at the detector arises purely from higher-order waveform modes. Since such modes are not accurately
described by the precessing IMRPhenom model but are contained in our \ac{NR} signals, we attribute the observed bias and reduction in recovered \SNR to the incompleteness of
the model.

To confirm this, we have injected signals
generated with the precessing IMRPhenom model, and performed \PE recovery with this
same model. This test shows no appreciable biases in the recovered parameters. This suggests that in these cases,
other possible sources of bias (for example, due to the choice of priors in the Bayesian
analysis) did not have a significant impact on the results. Another possibility is that
the biases are caused by inaccuracies in the NR waveforms, but since we see similar
effects between waveforms calculated from both the BAM code and {\tt SpEC}, we consider this unlikely.
We therefore conclude that the \PE biases for the configurations with $\theta_\mathrm{JN} \approx 90\degree$ in
Fig.~\ref{fig:phenp_posteriors} indeed arise from a lack of fidelity between the waveform model
and the full \NR signals.

With this in mind, a practical question becomes \textit{what fraction of
future detections will incur such biases?} Unfortunately, without knowledge of
the mass distribution of future observations, and given the small sample of
configurations analyzed here, we cannot answer this question in full
generality.

However, our investigation demonstrates that large parameter biases occur only
in strongly inclined binaries.
For these orientations, the observed \GW signal is weaker than for
other orientations, which significantly reduces their detectability.
As an illustration, we can estimate that
only 0.3\% of observable sources will fall into a $30^\circ \times 30^\circ$
region in inclination and polarization around the point of minimal amplitude
(which we take approximately as the point of maximal bias). Details of this
calculation are presented in Appendix~\ref{sec:angle_distribution}.

\subsection{Higher modes}
\label{sub:higher_modes}

\IMR waveform models for spinning \BBH with
higher modes are not yet available. Our analysis uses the precessing IMRPhenom
waveform model which includes only spherical harmonics of multipole $\ell=2$. In
Secs.~\ref{sub:aligned_spin} and~\ref{sub:precession} we have analyzed parameter
recovery of complete NR signals containing higher modes up to multipole $\ell=8$,
although, since the GW frequency scales with $m$, harmonics with $m \ge 3$ turn
on within the detector band because of the limited length of the \NR waveforms.

Higher modes are likely unimportant for nearly equal-mass systems and
become more relevant as the mass ratio
decreases~\cite{Pekowsky:2012sr,Varma:2014jxa,Bustillo:2015qty,pan:2011gk,Capano:2013raa}.
The importance of higher modes also increases with the total mass of
the system as the merger part of the signal moves into the most
sensitive part of the \aLIGO band.
Because no recovery waveform families exist which incorporate higher modes,
we test their importance by changing the injected waveform: Starting from the
precessing simulation \texttt{SXS:BBH:0308} (see Table~\ref{tbl:NRparams}),
we inject (a) the full \NR waveform with all modes up to $\ell=8$. And (b)
a ``truncated'' \NR waveform that consists only of the $\ell=2$ modes.
All injections are recovered with precessing IMRPhenom templates.

\begin{figure}[htbp]
  \centering
  \includegraphics[width=.48\textwidth,clip=true, trim=0 20 20 20]{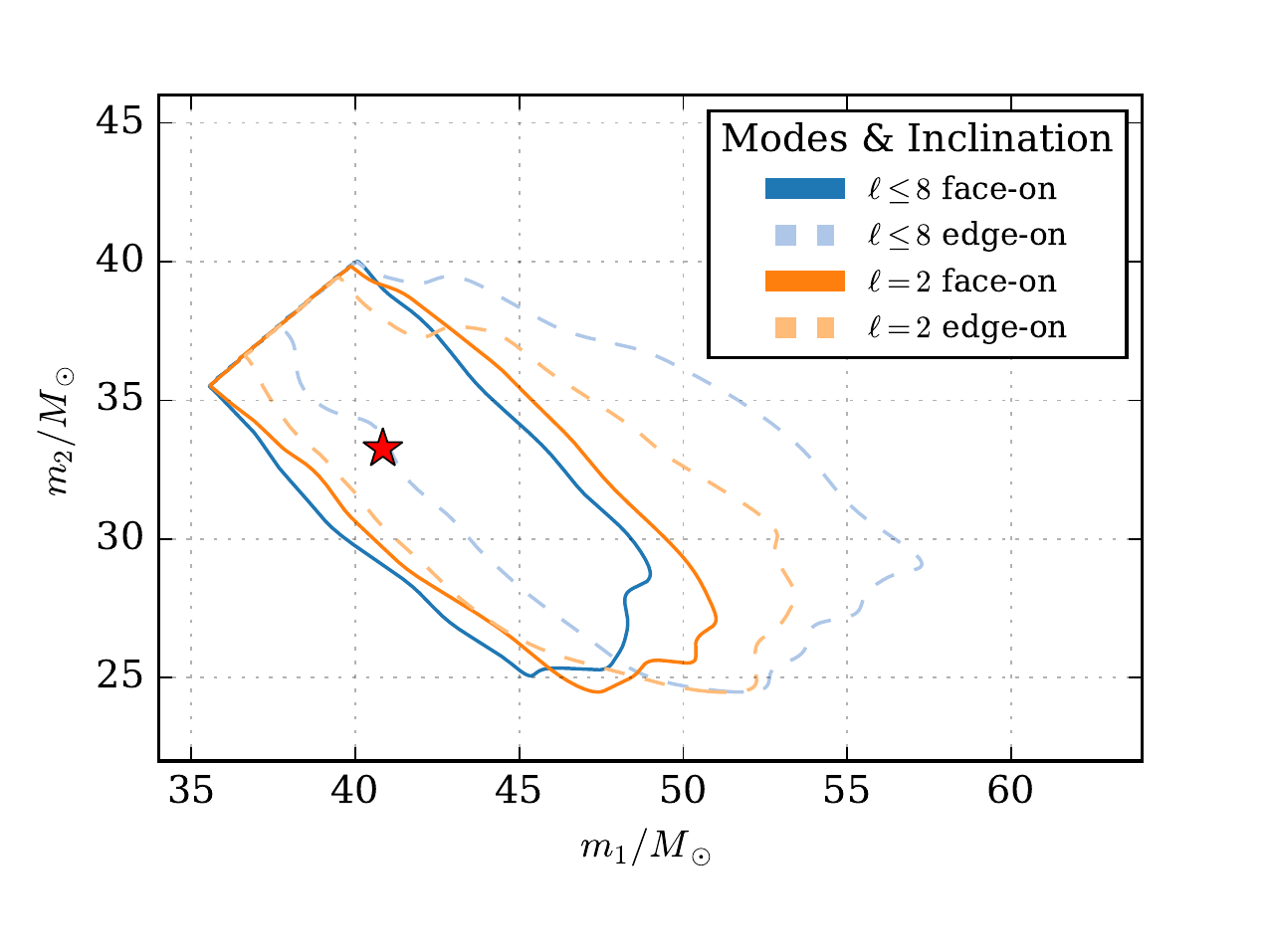}
    \includegraphics[width=.48\textwidth,clip=true, trim=0 20 20 20]{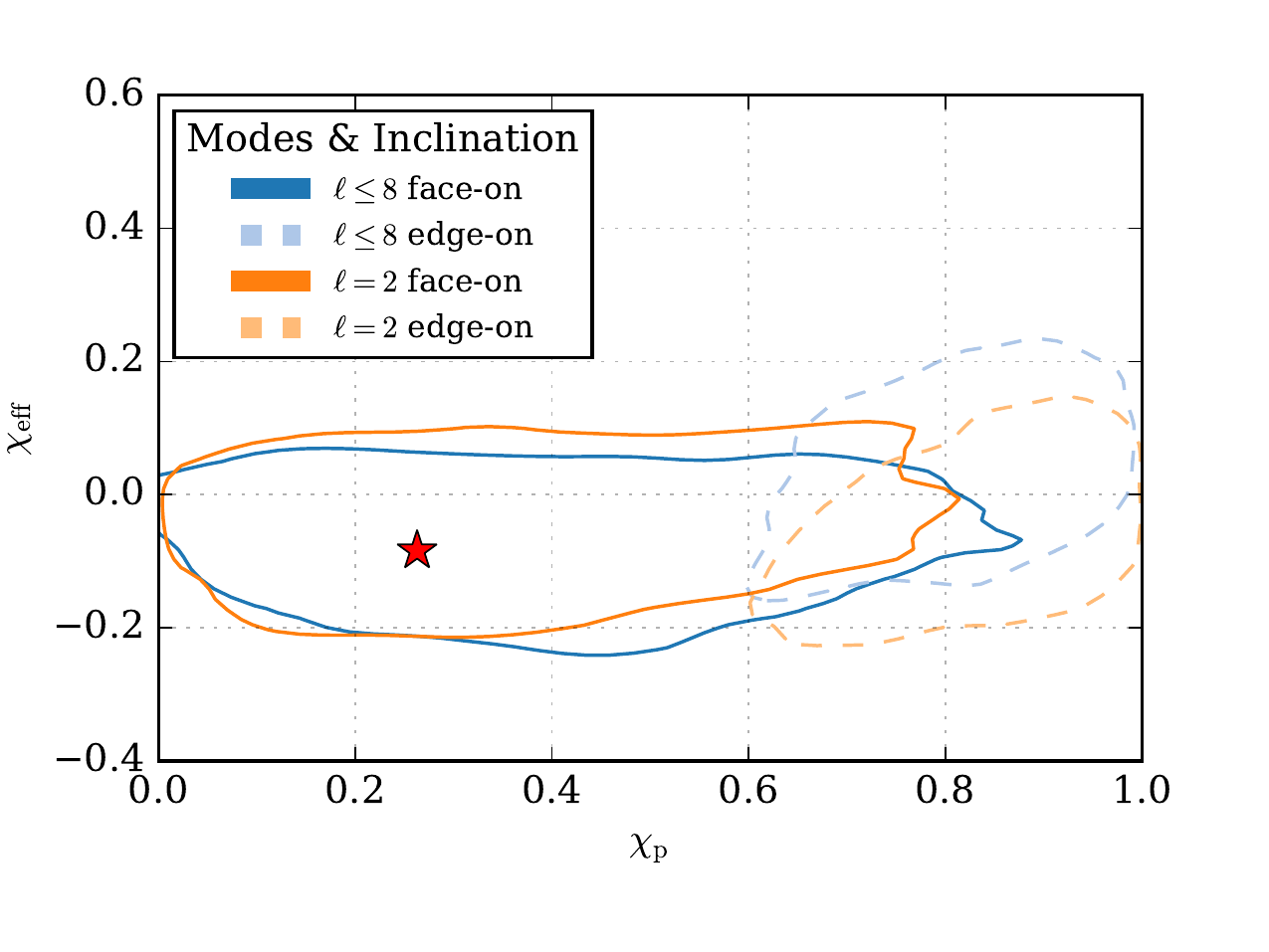}
  \caption{
    Results for precessing NR injections (\texttt{SXS:BBH:0308}) with face-on or edge-on inclination ($\theta_\mathrm{JN} = 6\degree$ and $84\degree$, respectively) and either including higher harmonics up to $\ell=8$ (compare with Fig.~\ref{fig:phenp_posteriors}) or just the $\ell=2$ modes in the mock signal. All injections are performed at fiducial polarization angle $\psi=82\degree$. The precessing IMR-\\Phenom model is used as the template waveform. We show two-dimensional $90\%$ credible regions for component masses and effective spins.
  }
  \label{fig:plots_SXS_0234_IMRPhenomPv2_higher_modes}
\end{figure}

Our results are summarized in Fig.~\ref{fig:plots_SXS_0234_IMRPhenomPv2_higher_modes}.
When the binary is
viewed face-on there is very good agreement in the posteriors irrespective of
whether the \NR mock signal includes all higher harmonics or just the $\ell=2$ modes
and the posteriors peak close to the true parameter values. However, when viewed
from edge-on inclination, parameter recovery is biased.
The larger mass $m_1$ is somewhat overestimated, and the effective
precession spin parameter is significantly overestimated, indicating erroneously
a nearly maximally precessing system, with the actual injected $\chi_p$
far outside the recovered $90\%$ credible region. These biases arise for both the full
\NR waveform, $\ell \leq 8$, and ``truncated'' \NR waveform, $\ell=2$.
At edge-on
inclination the higher harmonics contribute more to the GW signal and their
inclusion or absence also influences parameter recovery.
But this effect is much smaller than the bias arising from the inclination of the signal.

We can refine the conclusions of Sec.~\ref{sub:precession} and say that the
precessing IMRPhenom waveform model leads to biased parameter recovery for only
a very small fraction of orientations in the vicinity of \TheEvent{}, namely
when the system is viewed close to edge-on and if the GW polarization
happens to be unfavorable. For these exceptional cases we find that most of
the modeling error stems from the $\ell=2$ modes while neglecting modes with
$\ell>2$ in the model only causes additional small modeling errors.
See~\cite{Abbott:2016apu} for a more detailed discussion of cases where
higher modes can provide additional information.

To further quantify the effect of higher modes, we compute the
mismatch between the NR waveforms including only
$\ell=2$ modes and those including all modes up to $\ell=8$. The
computation is done in the same way as
in Tables~\ref{tbl:NRparams} and~\ref{tbl:NRparams_additional} using the highest
available resolution. We find that the typical mismatches for
configurations with mass-ratios and spins compatible with \TheEvent{}
(such as \texttt{SXS:BBH:0308, SXS:BBH:310}) are of order few $\times
10^{-3}$, rising to few $\times 10^{-2}$ for configurations with high
spin (e.g. \texttt{SXS:BBH:0233, SXS:BBH:0257, SXS:BBH:0531}) and
become the largest ($\sim0.1$) for mass ratio $q=0.125$
(\texttt{SXS:BBH:0065}). The mismatches are also higher for higher
inclinations, becoming largest for edge-on configurations.

The mismatch at high mass ratio is considerably larger than the
fiducial limit from a Fisher argument, consistent with previous
studies that found that subdominant modes become increasingly
important for higher mass ratio~\cite{Pekowsky:2012sr,Capano:2013raa,
  Healy:2013jza, Varma:2014jxa,Bustillo:2015qty}. The Fisher matrix
criterion is conservative, and violating it means that explicit PE
studies must be performed to assess the effects of neglecting higher
modes (see Appendix
\ref{sec:parameter_estimation_results_for_additional_nr_configurations}
for additional NR injections using the fiducial extrinsic
parameters). While we do not find significant biases for the cases
and parameters considered in this work, we expect that higher modes
will become important with larger inclinations and mass ratios.

\subsection{Eccentricity}
\label{sub:eccentricity}

Since \IMR waveform models including spin and
eccentricity are not currently available, we assess the effect of
eccentricity on \PE by injecting \NR waveforms of varying eccentricity
and studying \PE using a non-eccentric waveform model.

We use a family of \NR waveforms produced with {\tt SpEC} with
spins aligned with the orbital angular momentum with mass-ratio $q = 0.82$,
and aligned component spins $\chi_{1L} = 0.33$, $\chi_{2L} = -0.44$,
a configuration comparable to the parameters of \TheEvent.  The waveforms
in this family vary in their orbital eccentricity, cf. Table~\ref{tbl:NRparams}.

There is no unambiguous \GR definition of eccentricity,
so we calculate an eccentricity estimator~\cite{Mroue:2010re} from the
instantaneous frequency of the \GW using a Newtonian model.
We assume that the \GW frequency is twice the orbital
frequency of a Newtonian orbit, but fit for additional degrees of freedom
to model \GR effects such as inspiral and precession of the orbit.

We estimate the eccentricity by fitting a short portion of the
instantaneous \GW frequency, $\omega_\mathrm{GW}$, to
the form
\begin{align}\label{eqn:omega-GW-eccentric}
  \omega_\mathrm{GW} &= 2 n(t) \frac{\sqrt{1-e^2}}{\bigl[1-e \cos\left(u(t)\right) \bigr]^2} \\
\label{eqn:meanmotion}
  n(t) &= n_0 \left [1+a(t-t_\mathrm{ref}) \right ] \\
\label{eqn:u-eccentric}
  u(t) &= 2\pi(t-t_0)/P
\end{align}
in a neighbourhood of a reference time $t_\mathrm{ref}$. This form for $\omega_\mathrm{GW}$
is twice the orbital angular frequency expected from a Newtonian
eccentric orbit, with the slow inspiral modeled as a linear variation
of the parameter $n$ with time. $t_0$ is a fitted parameter representing the time
of pericentre passage, and a local maximum in $\omega_\mathrm{GW}$. 
We do not enforce the Newtonian
relation $n = 2 \pi / P$, since it is broken in the \GR case
by pericentre advance.  $u(t)$ would properly be obtained using
the Kepler equation. However, we do not find this necessary, and have
effectively expanded it in small $e$. This expansion leads to good fits for
the small values of $e$ that we are simulating. It is
necessary to include the nonlinear terms in $e$ for the large-scale
behavior of $\omega_\mathrm{GW}$ in order to get a good fit when
$e \gtrsim 0.1$.  We find that using the coordinates of the horizon
centroids, instead of the GW frequency, leads to qualitative disagreement
with this simple Newtonian model, whereas the GW frequency matches very well.

Unlike the spin magnitudes and mass ratio, the eccentricity evolves
significantly in the $14$ orbits covered by the eccentric simulations, so
assigning a single number to each configuration requires selecting a
specific point in the evolution at which to quote the eccentricity.

We quote the eccentricity at a reference time $t_\mathrm{ref}$ at
which the mean \GW frequency $2 n$ is 23.8~Hz assuming
the source mass is 74~$M_{\odot}$.  This is $2M n =
0.0545424$ in geometric units.

We obtain eccentricities up to $e = 0.13$ at the reference time; see Table \ref{tbl:NRparams}.
Even ``circular'' \NR waveforms have a small eccentricity,
as it is not possible to reduce this to zero.  For example, the
smallest eccentricity in the family of waveforms considered here is
$\sim 10^{-4}$, not 0.

\begin{figure*}[htbp]
  \centering
\includegraphics[scale=0.6,trim=10 13 10 12,clip=true]{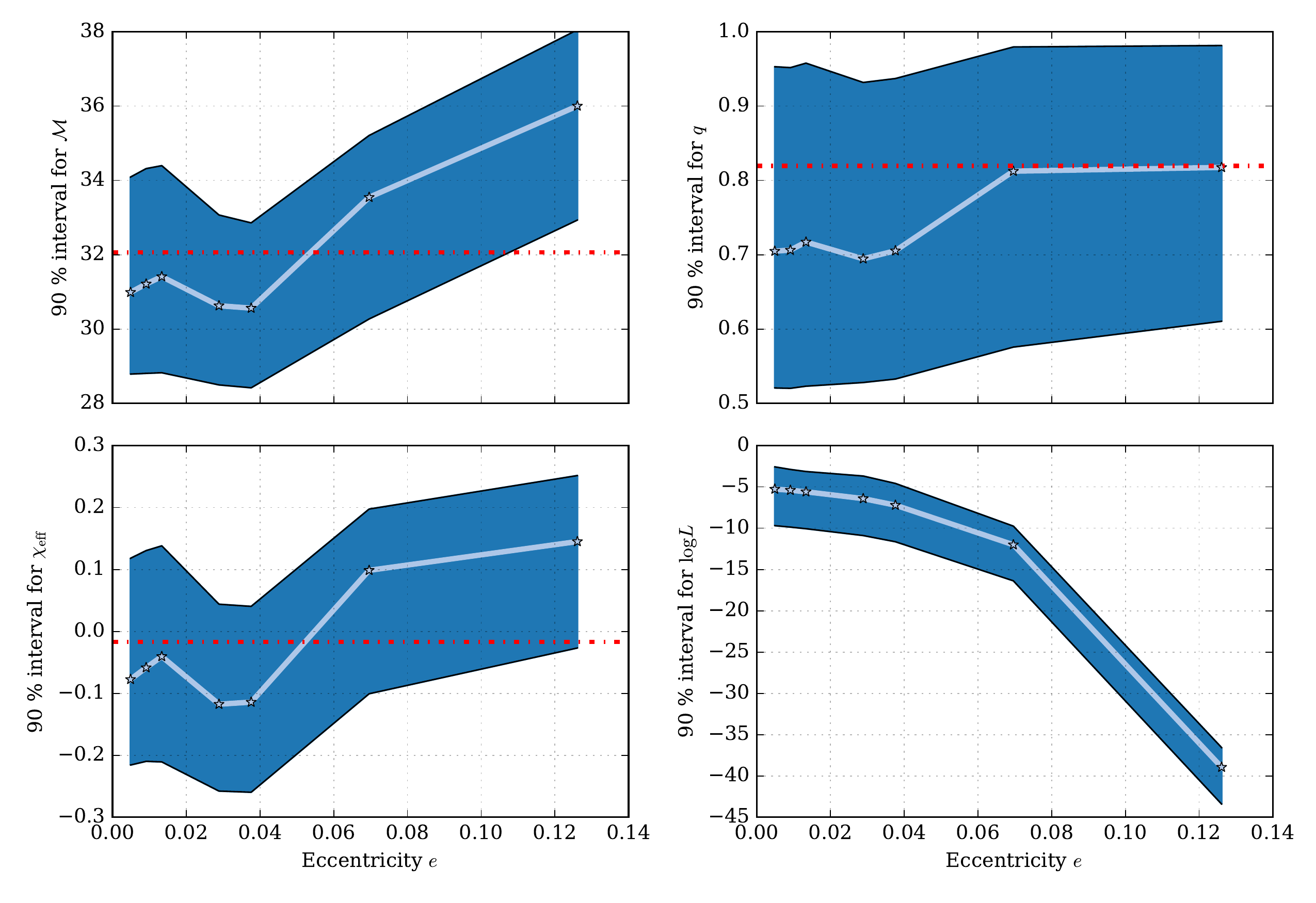}
  \caption{Parameter recovery of eccentric \NR mock signals with non-eccentric (quasi-circular)
  waveform templates. Shown are $90\%$ credible intervals of non-precessing EOBNR posteriors for
  \NR signals consistent with \TheEvent{} injected in zero-noise. Shown is the chirp mass, mass-ratio, effective aligned spin and log-likelihood. As the eccentricity of the mock signals increases the deviation of the median of the chirp mass from the injected value grows. There is no significant disagreement in the mass-ratio and aligned spin. The likelihood drops sharply as the eccentricity grows beyond $\sim 0.05$.}
  \label{fig:eccentric}
\end{figure*}

We inject the above eccentric aligned-spin \NR waveforms into zero noise and recover with the quasi-circular non-precessing EOBNR templates.
Fig.~\ref{fig:eccentric} shows posteriors for the chirp mass, mass-ratio and aligned spin on the larger BH as a function of eccentricity.
We find that eccentricities smaller than $\sim 0.05$ in the injected \NR waveform (with the eccentricity definition introduced above) do not strongly affect parameter recovery and lead to results comparable to quasi-circular NR waveforms. Biases occur for larger eccentricity.
The right panel of Fig.~\ref{fig:eccentric} shows how the log likelihood drops sharply if the eccentricity is above $0.05$ and the disagreement between the eccentric signal and quasi-circular template increases.

\subsection{Effect of detector noise}
\label{sub:effect_of_detector_noise}

So far in this study we have focussed on \NR injections in zero noise
using only an estimated \ac{PSD} from the detectors in order to assess
waveform systematics.
The results obtained with this method are missing two potentially important effects:
\begin{itemize}
\item [$\circ$] While we obtain the posterior probability density function effectively averaged
over many noise realizations, the zero-noise method does not assess how noise realizations
with typical deviations from the average will affect the posteriors.
\item [$\circ$] The usual interpretation of our credible intervals relies on the assumption that both
our signal and noise model are an appropriate description of the data. The previous sections addressed the signal model, but the zero-noise method does not take into account the
properties of actual detector noise, such as non-Gaussianity, non-stationarity and
inaccuracies in \ac{PSD} estimations.
\end{itemize}

In this section we study the variability of the posteriors for a selected
NR waveform \texttt{SXS:BBH:0308} for several noise realizations. We
compare with those examples the total uncertainty of \PE (including noise realization)
to the systematic error due to waveform model uncertainty from the previous sections.

We use LIGO-Hanford and LIGO-Livingston data from Monday September 14, 2015, 
surrounding~\TheEvent{}. This data is produced using an updated calibration model, 
as described in~\cite{Tuyenbayev:2016xey,Karki:2016pht}, which gives smaller 
uncertainties than the original $10\%$ in amplitude and $10 \degree$ in phase~\cite{Abbott:2016jsd} used for the first results~\cite{Abbott:2016blz, PhysRevLett.116.241102}.
For~\TheEvent{}, the standard deviations of the prior distributions for the amplitude 
and phase uncertainty due to calibration are (as in Table III of~\cite{TheLIGOScientific:2016pea}): amplitude Hanford: $4.8\%$, Livingston: $8.2\%$
and phase Hanford: $3.2 \degree$, Livingston: $4.2 \degree$.
The PE runs marginalize over calibration uncertainties 
with a spline model~\cite{SplineCalMarg,PhysRevLett.116.241102}.

\begin{figure}[htbp]
  \centering
   \includegraphics[width=.48\textwidth,clip=true, trim=10 1 20 22]{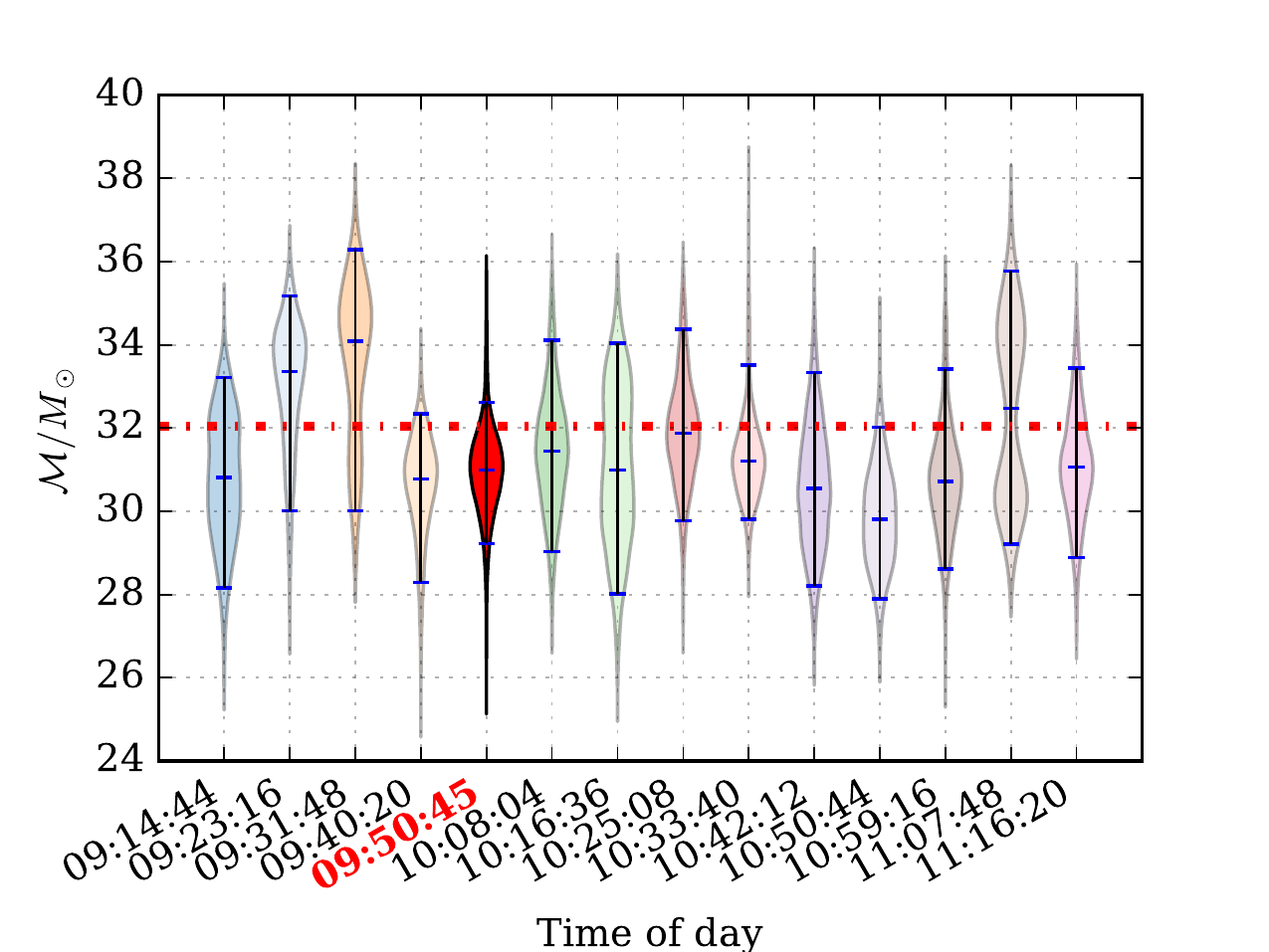}
  \includegraphics[width=.48\textwidth,clip=true, trim=10 1 20 22]{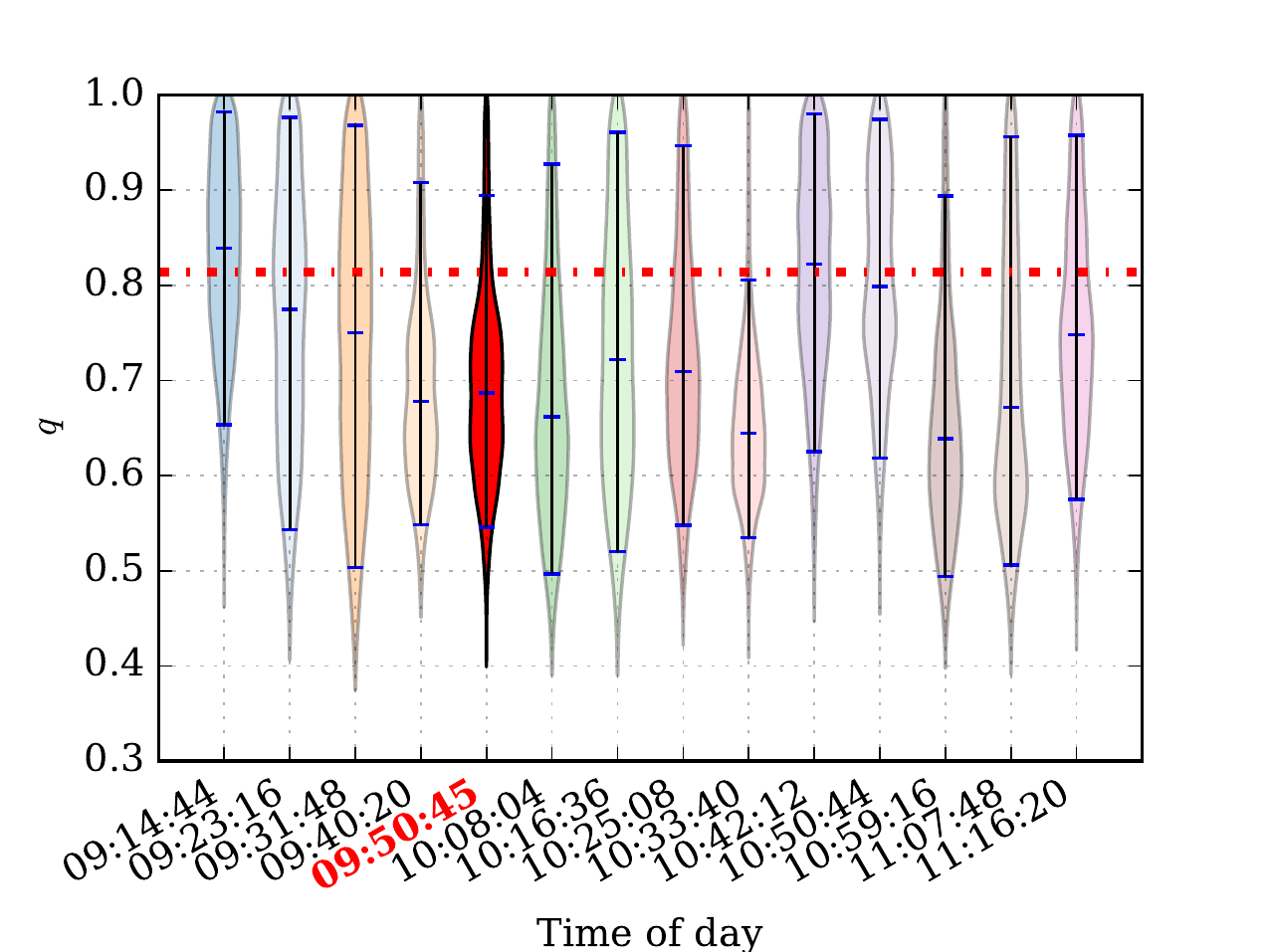}
  \caption{
    Violin plots of the posteriors for an NR waveform (\texttt{SXS:BBH:0308}) injected into
    13 different noise realizations on Monday September 14, 2015 (times in GMT) in the LIGO-Hanford and LIGO-Livingston data
    surrounding \TheEvent{}. The signal uses the fiducial values for the inclination and polarization angles.
  The \ac{PE} runs used the non-precessing EOBNR as a template and spline calibration marginalization.
  The violins extend over the entire range of the \PDF with medians and $90\%$ credible
  intervals indicated by lines.
  The injected values of the parameters are marked by a red dash-dotted line.
  We show a violin for parameter recovery in zero-noise in red, labeled with the time of \TheEvent{}.
    }
  \label{fig:C02_injections_0234_ROM_violin}
\end{figure}

We expect that for 90\% of the noise realizations the 90\% credible interval contains
the injected value for a given quantity when both our model of the waveform
and our noise-model (including the \ac{PSD}) are correct. We see from
Fig.~\ref{fig:C02_injections_0234_ROM_violin}
that most of the posteriors for the 13 different noise realizations agree
reasonably, except for \texttt{11:07:48} where the posterior is bimodal in the chirp
mass and thus very different from the zero-noise posterior.
Results are broadly consistent between the
EOBNR and IMRPhenom waveform models, and we find no
evidence that the assumptions motivating our zero-noise study are violated.

\subsection{Effect of numerical errors}
\label{sub:effect_of_numerical_errors}

\NR simulations of black hole binaries can only be carried out with
finite numerical resolution, which gives rise to truncation
errors. {\tt SpEC} uses \emph{hp} adaptive mesh refinement to ensure
accuracy and efficiency~\cite{Szilagyi:2014fna}.
Each numerical resolution is indexed by ${\rm Lev}=0,\dots,n$.
The truncation error is estimated at every step during the evolution, and the number
of basis functions (or, equivalently, collocation points) is adjusted to ensure that
the truncation error in all subdomains is less than a desired
threshold (see e.g.~\cite{Szilagyi:2014fna} for more details).

Another possible source of error is in the extraction of the gravitational waveform
itself. In {\tt SpEC}, GWs are computed using the standard Regge-Wheeler-Zerilli (RWZ)
formalism~\cite{PhysRev.108.1063,PhysRevLett.24.737,Sarbach:2001qq}. The
waveforms are extracted on a sequence of concentric coordinate
spherical shells centered on the origin of the grid
\cite{0264-9381-26-7-075009}. To mitigate gauge and finite radius
effects, the gravitational waveforms are then extrapolated to null
infinity by performing polynomial fitting in powers of $1/r$
\cite{Boyle:2009vi}. We label the polynomial degree of the fit by
$N=2,3,4$.

To assess the overall error, we choose a representative configuration
\texttt{SXS:BBH:0308} consistent with \TheEvent{} at fiducial orientation
(see Table~\ref{tbl:NRparams}) and compare the posteriors for different
numerical grid resolutions and extrapolation orders.

\begin{figure}
  \centering
    \includegraphics[width=.48\textwidth,clip=true, trim=10 20 20 20]{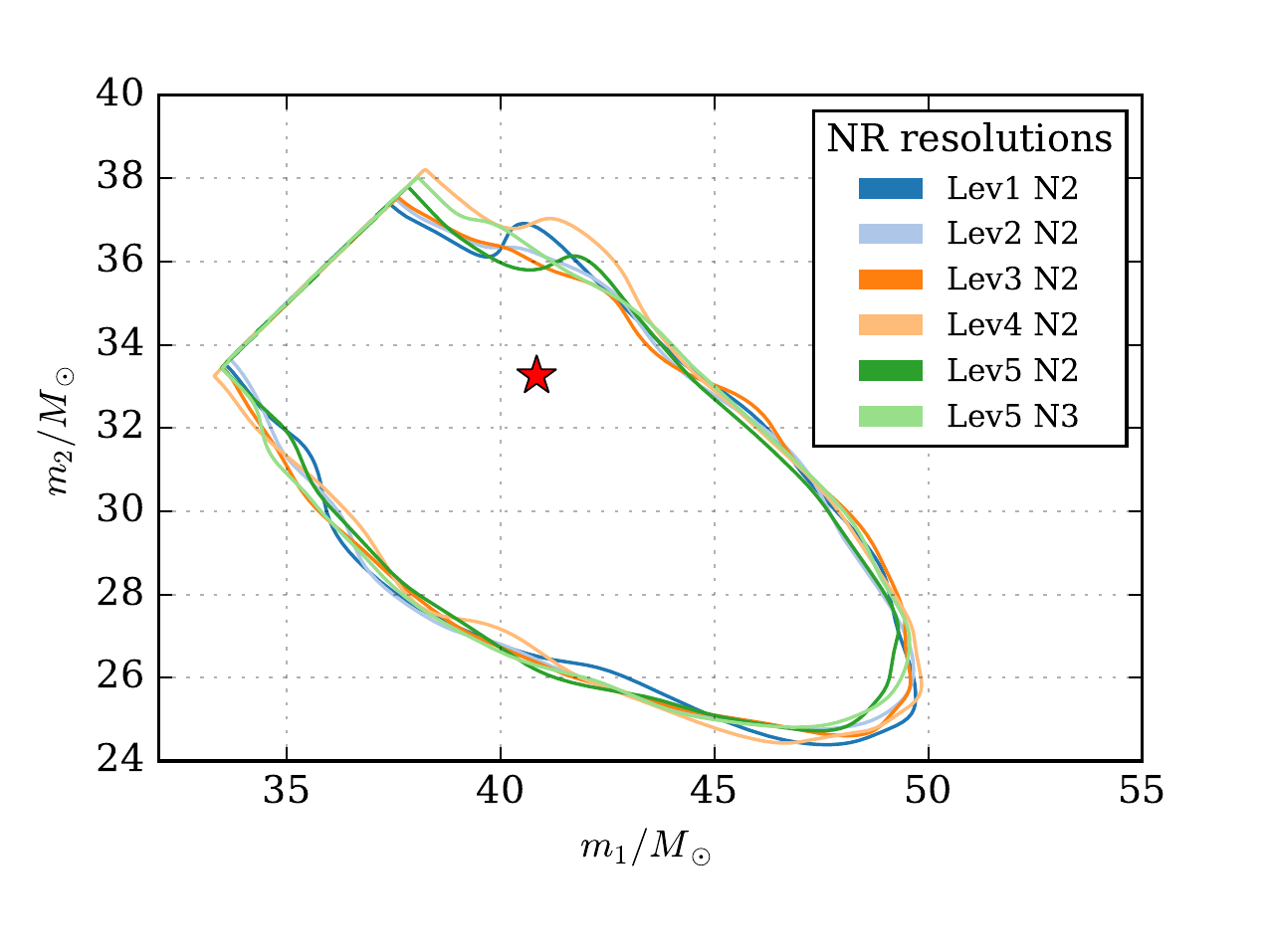}
    \includegraphics[width=.48\textwidth,clip=true, trim=10 20 20 20]{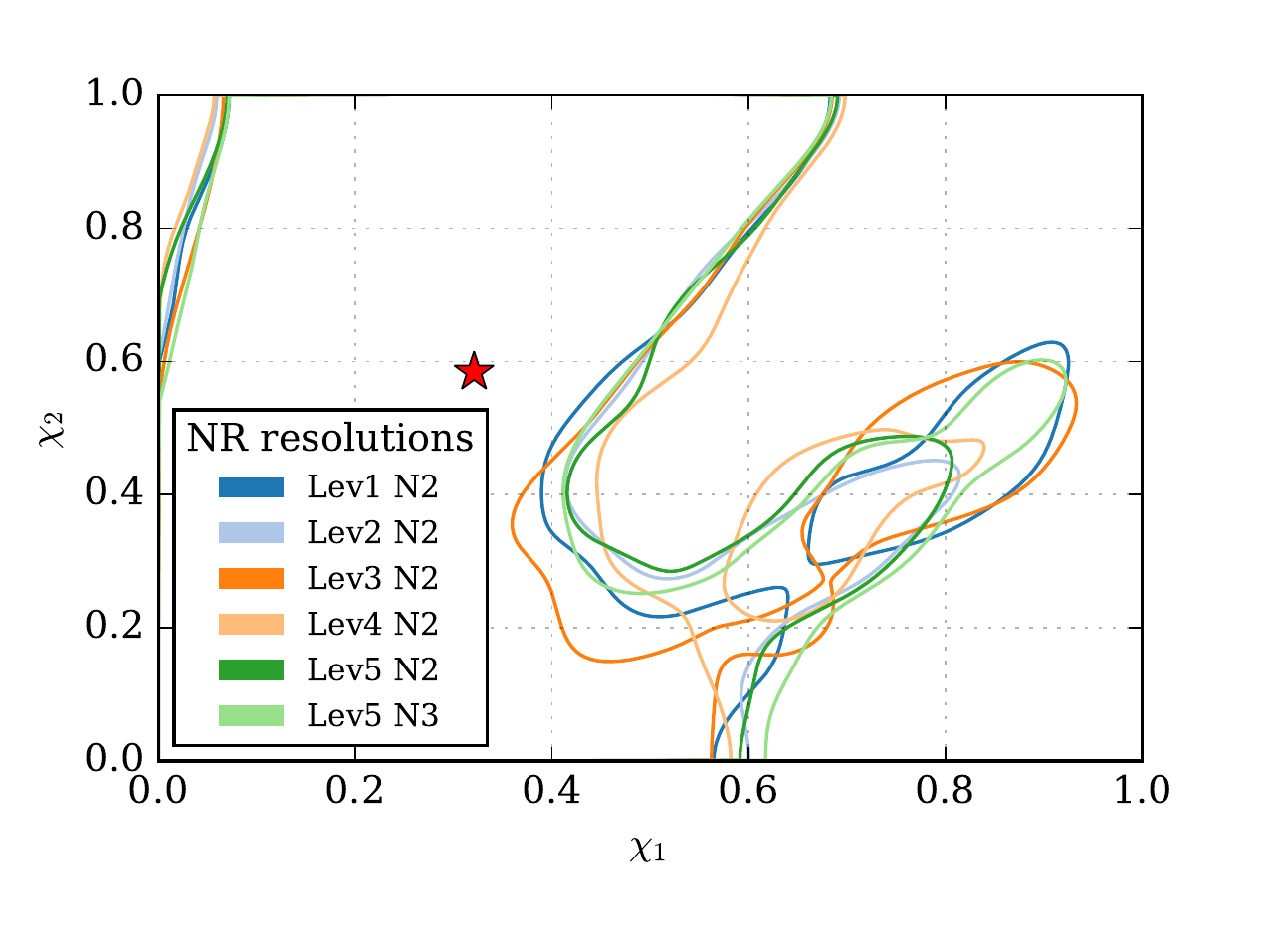}
  \caption{$90\%$ credible regions for configuration \texttt{SXS:BBH:0308} with numerical resolutions $\mathrm{Lev}=\{1,2,3,4,5\}$ and extrapolation orders $\mathrm{N}=\{2,3\}$.
  We find consistent results between the different resolutions and extrapolation orders, indicating that numerical errors are a negligible source of parameter biases.
  }
  \label{fig:NR_resolutions_IMRPhenomPv2}
\end{figure}

In Fig.~\ref{fig:NR_resolutions_IMRPhenomPv2} we show kernel density estimates of $90\%$ credible regions for posteriors from \ac{PE} simulations on this \ac{NR} waveform in zero noise varying the resolution and the extrapolation order. We find that the results for this \NR waveform with different resolutions and extrapolation orders agree extremely well.

We expect these results to be typical for all of the \NR waveforms that we have used. The
mismatch error of the BAM waveforms (see Table~\ref{tbl:NRparams}) is comparable to that of the representative
SXS configuration \texttt{SXS:BBH:0308}, which suggests that parameter biases due to numerical error will
also be negligible. We therefore conclude that our results are robust and the numerical error is not the dominant error source.

\section{Discussion}
\label{sec:discussion}

The parameters of the source of the first \GW observation,
\TheEvent{}, were analyzed using waveform models from
non-precessing~\cite{Taracchini:2013rva,Purrer:2015tud} and
precessing~\cite{Hannam:2013oca,Pan:2013rra,Babak:2016tgq} \BBH coalescences.
Both waveform models
were calibrated to \NR simulations, and are expected to be reliable for binary
configurations similar to \TheEvent{}, i.e., black holes with comparable masses
and low spins. The three models gave
consistent results~\cite{TheLIGOScientific:2016pea,PhysRevLett.116.241102,Abbott:2016izl,TheLIGOScientific:2016pea}.
Nonetheless, there are several possible sources of systematic errors:
the precessing IMRPhenom and EOBNR models were calibrated only to non-precessing \NR simulations~\cite{Husa:2015iqa,Khan:2015jqa,
Pan:2013rra,Babak:2016tgq}. In the case of the IMRPhenom model,
the precession effects are described with approximate \PN expressions;
the six-dimensional spin-parameter space is described using only three
judiciously chosen parameters, which were motivated by the dominant aligned
and precession spin effects during the inspiral; and both models include only
partial information about the sub-dominant harmonic modes of the signal.

The present study expands on a brief analysis of parameter biases
reported in Refs.~\cite{Abbott:2016blz,PhysRevLett.116.241102,Abbott:2016izl} which indicated
that the various waveform model deficiencies do not significantly
impact \PE for \TheEvent{}. Here, we use waveforms obtained
by direct numerical solutions of the full Einstein equations, and
inject these as mock signals into simulated \aLIGO detector noise.
Because of the high quality of numerical solutions of Einstein's
equations (cf. Fig.~\ref{fig:NR_resolutions_IMRPhenomPv2}), the
numerical waveforms can be taken as the prediction of Einstein's
equations, with negligible deviations from the exact \GR waveforms.
We then perform \PE studies with the waveform models
employed in~\cite{PhysRevLett.116.241102} and carefully document any deviation of
the recovered parameters from the parameters of the numerically
simulated \BBH systems.

The present study focuses specifically on BBH parameters comparable to
those of \TheEvent{}.  This is a fairly high-mass, nearly equal mass BBH
system, with nearly vanishing effective spin parallel to the orbital angular momentum,
$\chi_{\rm eff}$.  The spin-magnitudes and spin-directions are not significantly constrained,
except for the measurement of $\chi_{\rm eff}$.

The first study reported here concerns aligned-spin BBH systems, which
do not precess.  Recovering the parameters of such an injection with
non-precessing EOBNR and IMRPhenom waveform models yields unbiased
recovered parameters, where the uncertainty in the recovered
parameters is dominated by statistical errors, without noticeable
systematic biases (cf. Fig.~\ref{fig:plots_SXS_Ossokine_0233_HM}).
This result is consistent with extensive previous studies of
aligned-spin waveform models, e.g.~\cite{aasi:2014tra,Kumar:2016dhh,hinder:2013oqa}.

Focusing on the less-studied case of precessing binaries,
Figs.~\ref{fig:plots_mass_NR_SO_prec}
and~\ref{fig:plots_spins_NR_SO_prec} demonstrate that a precessing NR
waveform near fiducial parameters consistent with \TheEvent{} is recovered very well.
The parameters of the injected NR waveform are near the centers of the
recovered 90\% credible intervals in the noise-free case,
demonstrating confidently that any biases in the waveform models are small.
For the spin-recovery,
$\chi_{\rm eff}$ is measured comparably well as in the \TheEvent{}
\PE studies~\cite{PhysRevLett.116.241102,Abbott:2016izl}; the
precession effects encoded in $\chi_p$ are not meaningfully constrained,
again consistent with~\cite{PhysRevLett.116.241102,Abbott:2016izl}.  The most likely
parameters of \TheEvent{} suggest that the binary's inclination is nearly
face-off, a configuration for which
precession-induced modulations of the waveforms are small.
Fig.~\ref{fig:phenp_posteriors} presents a study of different
angles $\theta_\mathrm{JN}$ between the total angular momentum and the line-of-sight,
based on two numerical waveforms
with different magnitude of precession spins $\chi_p$. For most
values of $\theta_\mathrm{JN}$, the injected parameters are recovered very well
with no apparent systematic bias. However, if the system is viewed
edge-on ($\theta_\mathrm{JN}\approx 90^\circ$) \emph{and} if the \GW
polarization happens to be unfavorable, biases can
arise. For these particular cases, the waveform amplitudes are
significantly diminished because the detector orientation is near a
node of \emph{both} waveform polarizations\footnote{The
injections of Fig.~\ref{fig:phenp_posteriors} are performed at fixed
signal-to-noise ratio, and so the diminished GW amplitude manifests itself in
a smaller injected distance, cf. lowest panels of
Fig.~\ref{fig:phenp_posteriors}.}, cf. Fig.~\ref{fig:td_response}. In these
cases, the waveforms can significantly differ from the shapes of the model
signal. Because only the dominant harmonic effects are included in the
precessing IMRPhenom waveform model, and the precession effects through merger are
captured only approximately, it is not surprising that parameter recovery is
biased in this regime.
As outlined in Appendix~\ref{sec:angle_distribution}, only a small fraction $\sim 0.3\%$ of BBH binaries detectable by \aLIGO in the vicinity of GW150914 will fall into this biased edge-on regime.

Conversely, given the wide posteriors of \TheEvent{},
there is a small chance that the system's parameters are significantly
different from the most likely values, and indeed, the source of \TheEvent{} could be
oriented edge-on. However, the posterior probability for strongly
precessing systems oriented in this way is so small, that the waveform
inaccuracies indicated by the study performed here (see
Sec.~\ref{sec:effects_of_pol} and Appendix~\ref{sec:angle_distribution})
should not significantly affect the 90\% credible
levels reported in the analysis of \TheEvent{}~\cite{PhysRevLett.116.241102}.

The studies presented in Figs.~\ref{fig:plots_SXS_Ossokine_0233_HM}
to~\ref{fig:phenp_posteriors} always inject the complete NR waveform (using
all available numerical $(\ell, m)$ modes), whereas the recovery was
performed with waveform models that model correctly only the $(2,\pm
2)$ modes.  Therefore, the lack of bias in
Figs.~\ref{fig:plots_SXS_Ossokine_0233_HM}
to~\ref{fig:phenp_posteriors} already indicates that for \TheEvent{}-like
signals, modeling of sub-dominant waveform modes is not necessary.
This is confirmed by the study reported in
Fig.~\ref{fig:plots_SXS_0234_IMRPhenomPv2_higher_modes}.

All waveform models used to analyze \TheEvent{} assume circular orbits. Vanishingly
small eccentricity is expected for field binaries~\cite{peters:1963ux,peters:1964},
and with the \GW observations, this assumption can now be confronted
with data. A measurement of eccentricity requires eccentric \IMR
waveform models, which are currently not available. But see~\cite{Huerta:2016rwp}.
However, we can inject eccentric mock NR waveforms, and quantify
\PE biases. This study is reported in Fig.~\ref{fig:eccentric}:
Eccentricities up to a few percent (measured at a \GW frequency of 25~Hz) do not
result in systematic \PE biases, and only marginally reduce the
likelihood. However, for $e>0.05$, the likelihood drops significantly and
parameter biases become appreciable.

Finally, we investigate the variations in \PE estimates for
different noise realizations. Injecting the identical \NR signal into
the \aLIGO detector data at 13 different times around the time of
\TheEvent{},
we find no evidence that our zero-noise injection study is based on false assumptions.

Overall, our analysis finds no significant bias of the original
analysis of \TheEvent{}~\cite{PhysRevLett.116.241102}. \TheEvent{}
lies in a region of parameter space for which non-precessing binaries have
received detailed study~\cite{aasi:2014tra,Kumar:2016dhh,hinder:2013oqa,Taracchini:2013rva,Khan:2015jqa}
and reliable waveform models exist. The parameter
estimation results in Refs.~\cite{PhysRevLett.116.241102,Abbott:2016izl} and in
the present paper suggest that either \TheEvent{} did not include significant precession,
or that precession effects did not leave a strong imprint on the signal. This is
consistent with the high mass of the source, which causes only a few waveform cycles to be
in \aLIGO's frequency band, and the comparable masses of the two black holes. In such
systems we expect that precession effects are difficult to unambiguously distinguish
unless the binary has a large inclination with respect to the observer.

In the vicinity of GW150914 we would expect biases if the source had significantly
higher SNR than $25$. In the high \SNR regime the statistical errors decrease linearly
with the inverse of the \SNR. For the results shown in Figs.~\ref{fig:plots_SXS_Ossokine_0233_HM}
to~\ref{fig:phenp_posteriors} parameters would therefore start to become biased at
SNRs above $\sim 70 - 100$.

\TheEvent{} lies in
a region of parameter space which is fairly easy to model: The
small number of observable GW cycles combined with a
mass-ratio close to unity and modest spin magnitudes make this system easy
to study with numerical relativity.  Indeed simulations covering this part of \BBH parameter space have
been available for several years
(e.g.~\cite{ajith:2012tt,hinder:2013oqa}) and are incorporated
in current waveform models.  Moreover, several
properties of \TheEvent{} suppress the importance of sub-dominant
waveform modes and the importance of precession: Comparable mass,
moderate spins, short duration, near face-on orientation.
Waveform models are less mature for lower-mass
systems, higher mass ratios, and larger spin magnitudes, and in these systems
precession-induced waveform modulations may be easier to discern in the data.
Therefore, we recommend that this study is repeated for \BBH \GW observations in
other regions of parameter space.

\acknowledgments
The authors gratefully acknowledge the support of the United States
National Science Foundation (NSF) for the construction and operation of the
LIGO Laboratory and Advanced LIGO as well as the Science and Technology Facilities Council (STFC) of the
United Kingdom, the Max-Planck-Society (MPS), and the State of
Niedersachsen/Germany for support of the construction of Advanced LIGO 
and construction and operation of the GEO600 detector. 
Additional support for Advanced LIGO was provided by the Australian Research Council.
The authors gratefully acknowledge the Italian Istituto Nazionale di Fisica Nucleare (INFN),  
the French Centre National de la Recherche Scientifique (CNRS) and
the Foundation for Fundamental Research on Matter supported by the Netherlands Organisation for Scientific Research, 
for the construction and operation of the Virgo detector
and the creation and support  of the EGO consortium. 
The authors also gratefully acknowledge research support from these agencies as well as by 
the Council of Scientific and Industrial Research of India, 
Department of Science and Technology, India,
Science \& Engineering Research Board (SERB), India,
Ministry of Human Resource Development, India,
the Spanish Ministerio de Econom\'ia y Competitividad,
the Conselleria d'Economia i Competitivitat and Conselleria d'Educaci\'o, Cultura i Universitats of the Govern de les Illes Balears,
the National Science Centre of Poland,
the European Commission,
the Royal Society, 
the Scottish Funding Council, 
the Scottish Universities Physics Alliance, 
the Hungarian Scientific Research Fund (OTKA),
the Lyon Institute of Origins (LIO),
the National Research Foundation of Korea,
Industry Canada and the Province of Ontario through the Ministry of Economic Development and Innovation, 
the Natural Science and Engineering Research Council Canada,
Canadian Institute for Advanced Research,
the Brazilian Ministry of Science, Technology, and Innovation,
Funda\c{c}\~ao de Amparo \`a Pesquisa do Estado de S\~ao Paulo (FAPESP),
Russian Foundation for Basic Research,
the Leverhulme Trust, 
the Research Corporation, 
Ministry of Science and Technology (MOST), Taiwan
and
the Kavli Foundation.
The authors gratefully acknowledge the support of the NSF, STFC, MPS, INFN, CNRS and the
State of Niedersachsen/Germany for provision of computational resources.

CFUIB simulations were carried out on the UK DiRAC Datacentric
cluster. The SXS simulations were carried out on HPC resources
provided by Compute Canada, the Research Corporation, and California
State University Fullerton, on the San Diego Supercomputer Center's machine Comet
and on the AEI Datura cluster. We further acknowledge
support from the Research Corporation for Science Advancement, and the
Sherman Fairchild Foundation.


\appendix

\section{Parameter estimation results for additional NR configurations} \label{sec:parameter_estimation_results_for_additional_nr_configurations}

In addition to the \ac{NR} runs presented in Table~\ref{tbl:NRparams}, we have also analyzed a set of supplementary
aligned-spin and precessing binary configurations as listed in Table~\ref{tbl:NRparams_additional}. Tables ~\ref{tab:aligned_summary} and~\ref{tab:precessing_summary} list the
\PE results for these additional configurations.

The aligned-spin cases, which span a range of mass ratios and spin magnitudes, were injected at the fiducial inclination angle of $\iota = 163\degree$ and analyzed with the non-precessing
EOBNR~\cite{Taracchini:2013rva, Purrer:2015tud} and IMRPhenom~\cite{Khan:2015jqa} waveform models.
The \NR signal includes all higher modes. Differences to results for injections with just the $\ell=|m|=2$ modes are very small.
Overall, we find highly consistent results between the
two waveform families. This is particularly true for the equal-mass cases, where both models recover the effective
spin (see Eq.~(\ref{eq:chieff})) and the chirp mass accurately, with the most biased physical parameter being the mass ratio.
Although, for decreased mass ratios we find that the recovery of $q$ improves.
We note, however, that for the equal mass case the mass ratio lies
on the boundary of the physically allowed space, so the median estimate must always be biased.
The effective aligned spin $\chi_\mathrm{eff}$ is well recovered with biases smaller than 0.1 for all cases. In addition, we observe a weak correlation between the bias in the chirp mass and the bias in the effective spin.
For \texttt{SXS:BBH:0257} and \texttt{SXS:BBH:0233} we find noticeable biases in the mass-ratio recovery of EOBNR. In addition, for \texttt{SXS:BBH:0233} we find very broad PDFs in the chirp mass for both models, but more markedly for EOBNR.
The posteriors obtained from the non-precessing EOBNR for \texttt{SXS:BBH:0257} are bimodal in chirp-mass, mass-ratio and effective spin. This simulation was previously found to significantly disagree with EOBNR~\cite{Kumar:2016dhh}, having a mismatch of about $10\%$.
While these configurations have high component spins and lie at the edge of the calibration
ranges of EOBNR (both cases) and IMRPhenom (\texttt{SXS:BBH:0233}) where their accuracy may be
diminished, we emphasize that for the configurations near the most likely parameters of
\TheEvent{} both models recover the \NR parameters very accurately.
We do not show results for the anti-symmetric combination of aligned spins since it is in general poorly constrained~\cite{purrer:2015nkh}.

For the three additional precessing cases, which were analyzed only with the precessing IMRPhenom waveform model~\cite{Hannam:2013oca}, we find
qualitatively similar results, although we emphasize again that the small number of configurations does not allow to make global
statements. Nevertheless, out of the three cases we find that the mass ratio is determined best for the $q=0.333$ run. Similar to the
aligned-spin runs, we find that large biases in $\chi_\mathrm{eff}$ are correlated with large biases in the chirp mass.
For all except one configuration, we find that the precession spin $\chi_p$ is underestimated, with only minimal improvement when the inclination angle is changed from nearly face-off to edge-on
inclination. $\chi_\mathrm{eff}$ on the other hand is very well determined with a bias smaller than $0.1$. Only for edge-on inclination and the fiducial polarization value can the bias become large.

 \begin{table*}
 \begin{ruledtabular}
  \begin{tabular}{lccccccccc}
    ID & q & $\boldsymbol{\chi}_{1}$ & $\boldsymbol{\chi}_{2}$ & $\chi_{\rm eff}$ & $\chi_{p}$ & $M\Omega$ & $N_{orbits}$ & e & $1 - \mathcal{O}_{\rm res}$ \\
    \hline
\texttt{SXS:BBH:0211} & 1.0 & (0.0,0.0,-0.8997)  & (0.0,0.0,0.8998) & 0.0 & 0.0 & 0.014107 & 22.3 & 0.00026 & $4.1\times10^{-4}$ \\
\texttt{SXS:BBH:0213} & 1.0 & (0.0,0.0,-0.7998)  & (0.0,0.0,0.7999) & 0.0 & 0.0 & 0.014346 & 22.3 & 0.00014 & $5.2\times 10^{-4}$\\
\texttt{SXS:BBH:0180} & 1.0 & (0.0,0.0,0.0) & (0.0,0.0,0.0) &  0.0 & 0.0 & 0.01227 & 28.2 & 0.00005 & $5.5\times10^{-4}$   \\
\texttt{SXS:BBH:0219} & 1.0 & (0.0,0.0,-0.5)  & (0.0,0.0,0.8998) & 0.2 & 0.0 & 0.014836 & 22.4 & 0.00033 & $2.5\times 10^{-4}$\\
\texttt{SXS:BBH:0229} & 1.0 & (0.0,0.0,0.65)  & (0.0,0.0,0.25) &   0.45 & 0.0 & 0.014879 & 23.1 & 0.00031 & $1.8\times 10^{-4}$\\
\texttt{SXS:BBH:0231} & 1.0 & (0.0,0.0,0.8998)  & (0.0,0.0,0.0) & 0.4499 & 0.0 & 0.014874 & 23.1 &  $<1\times 10^{-4}$ &  $2.4\times 10^{-4}$ \\
\texttt{SXS:BBH:0152} & 1.0 & (0.0,0.0,0.6)  & (0.0,0.0,0.6) & 0.6 & 0.0 & 0.015529 & 22.6 & 0.00043 & $3\times 10^{-5}$\\
\texttt{SXS:BBH:0214} & 1.0 & (0.0,0.0,-0.6249)  & (0.0,0.0,-0.25) & -0.4375 & 0.0 & 0.012637 & 24.4 & 0.00019 & $1.1\times 10^{-4}$\\
\texttt{SXS:BBH:0311} & 0.84 & (0,0,0.4199) & (0,0,0.3800) & 0.4017 & 0.0 &0.017507 & 18.2 & $1.2 \times10^{-4}$ & $2.9\times 10^{-4}$ \\
\texttt{SXS:BBH:0310} & 0.82 & (0,0,0) & (0,0,0) & 0.0 & 0.0 &  0.018230 & 15.2 & $7.7\times10^{-4}$ & $4.1\times10^{-5}$ \\
\texttt{SXS:BBH:0309} &  0.82 & (0.0, 0.0, 0.3302) & (0.0, 0.0, -0.4398) & -0.0165  & 0.0 & 0.017942 & 15.7 & 0.02763 & $5.4\times10^{-4}$ \\
\texttt{SXS:BBH:0305} & 0.82 & (0,0,0.3301) & (0,0,-0.4399) & -0.0166 & 0.0 &  0.018208 & 15.2 & $2.5\times10^{-4}$ & $3\times 10^{-4}$\\
\texttt{SXS:BBH:0019} & 0.667 & (0.0,0.0,-0.4995)  & (0.0,0.0,0.4995) & -0.0999 & 0.0 & 0.014604 & 20.4 & $<7.6\times 10^{-5}$ & $5.5\times 10^{-4}$\\
\texttt{SXS:BBH:0239} & 0.5 & (0.0,0.0,-0.3713)  & (0.0,0.0,0.8497) & 0.0358 & 0.0 &  0.014782 & 22.2 & $<9.1\times 10^{-5}$ & $2.2\times 10^{-5}$\\
\texttt{SXS:BBH:0257} & 0.5 & (0.0,0.0,0.8498)  & (0.0,0.0,0.8498) &  0.8498 & 0.0 & 0.016332 & 24.8 & 0.00011 & $3\times 10^{-4}$\\
\texttt{SXS:BBH:0233} & 0.5 & (0.0,0.0,-0.8713)  & (0.0,0.0,0.8497) & -0.2976 & 0.0  & 0.014232 & 22.0 & 0.00006 & $3.2\times 10^{-4}$ \\
\texttt{SXS:BBH:0065} & 0.125 & (0.0,0.0,0.4996)  & (0.0,0.0,0.0) &  0.4441 & 0.0 &
0.018869 & 34.0 & 0.00374 & $7.8\times 10^{-4}$\\[8pt]
\texttt{SXS:BBH:0522} & 0.588 & (0.0787,0.5729,-0.5527)  & (-0.0509,0.033,-0.7974) & -0.655 & 0.5588 & 0.015463 & 16.7 & 0.0002321 & $9.4\times 10^{-4}$\\
\texttt{SXS:BBH:0531} & 0.588 & (-0.2992,0.4469,0.5925)  & (0.0787,0.0269,0.7954)  & 0.6601 & 0.5509 & 0.017381 & 21.4 & $<1.5\times 10^{-4}$ & $2.8\times 10^{-4}$\\
\texttt{SXS:BBH:0049} & 0.333 & (0.4941,0.0733,0.0011)  & (-0.0002,-0.008,0.4993) &
0.1267 & 0.4995 & 0.017518 & 19.4 & 0.00041 & n/a
  \end{tabular}
  \caption{Parameters of additional NR runs from the public SXS waveform catalog~\cite{Mroue:2013xna,SXSCatalog}, the non-public catalog~\cite{Chu:2015kft} and new simulations targeted at \TheEvent{}~\cite{Blackman:2017dfb}. The columns are as defined in Table~\ref{tbl:NRparams}.
   }
\label{tbl:NRparams_additional}
  \end{ruledtabular}
\end{table*}

\begin{table*}[p]
\centering
\begin{ruledtabular}
  \begin{tabular}{@{}lrrrrrrrrrrrrrrrr@{}}
& \multicolumn{4}{c}{$\mathcal{M}$} & \phantom{abcd}& \multicolumn{4}{c}{$q$} & \phantom{abcd}& \multicolumn{4}{c}{$\chi_\mathrm{eff}$}\\
{} &  True &  Median &  Bias &  90\% CI &        &  True &  Median &  Bias &
90\% CI &        &  True &  Median &  Bias &  90\% CI &        \\\hline
\texttt{SXS:BBH:0211}\\
EOBNR & 32.25 &   31.22 &  1.03 &     5.15 &  &  1.00 &    0.71 &  0.29 &
    0.42 &  &  0.00 &   -0.06 &  0.06 &     0.33 &  \\
IMRPhenom  & 32.25 &   31.81 &  0.44 &     4.23 &  &  1.00 &    0.82 &  0.18 &
    0.38 &  &  0.00 &   -0.02 &  0.02 &     0.24 &  \\[3pt]
\texttt{SXS:BBH:0213}\\
EOBNR & 32.25 &   31.04 &  1.21 &     5.70 &  &  1.00 &    0.68 &  0.32 &
    0.44 &  &  0.00 &   -0.07 &  0.07 &     0.35 &  \\
IMRPhenom  & 32.25 &   32.16 &  0.09 &     4.28 &  &  1.00 &    0.81 &  0.19 &
    0.40 &  &  0.00 &    0.00 & -0.00 &     0.24 &  \\[3pt]
\texttt{SXS:BBH:0180}\\
EOBNR & 32.25 &   31.20 &  1.06 &     5.34 &  &  1.00 &    0.71 &  0.29 &
    0.43 &  & -0.00 &   -0.07 &  0.07 &     0.33 &  \\
IMRPhenom  & 32.25 &   31.88 &  0.38 &     4.16 &  &  1.00 &    0.82 &  0.18 &
    0.38 &  & -0.00 &   -0.02 &  0.02 &     0.24 &  \\[3pt]
\texttt{SXS:BBH:0219}\\
EOBNR & 32.25 &   32.67 & -0.41 &     3.29 &  &  1.00 &    0.80 &  0.20 &
    0.40 &  &  0.20 &    0.21 & -0.01 &     0.21 &  \\
IMRPhenom  & 32.25 &   32.16 &  0.10 &     3.60 &  &  1.00 &    0.79 &  0.21 &
    0.40 &  &  0.20 &    0.20 &  0.00 &     0.21 &  \\[3pt]
\texttt{SXS:BBH:0229}\\
EOBNR & 32.25 &   32.41 & -0.16 &     3.94 &  &  1.00 &    0.80 &  0.20 &
    0.43 &  &  0.45 &    0.43 &  0.02 &     0.27 &  \\
IMRPhenom  & 32.25 &   31.92 &  0.33 &     2.94 &  &  1.00 &    0.79 &  0.21 &
    0.42 &  &  0.45 &    0.43 &  0.02 &     0.17 &  \\[3pt]
\texttt{SXS:BBH:0231}\\
EOBNR & 32.25 &   32.48 & -0.23 &     3.95 &  &  1.00 &    0.80 &  0.20 &
    0.43 &  &  0.45 &    0.43 &  0.02 &     0.28 &  \\
IMRPhenom  & 32.25 &   31.97 &  0.28 &     3.00 &  &  1.00 &    0.79 &  0.21 &
    0.44 &  &  0.45 &    0.43 &  0.02 &     0.18 &  \\[3pt]
\texttt{SXS:BBH:0152}\\
EOBNR & 32.25 &   32.85 & -0.59 &     3.63 &  &  1.00 &    0.81 &  0.19 &
    0.44 &  &  0.60 &    0.62 & -0.02 &     0.30 &  \\
IMRPhenom  & 32.25 &   31.79 &  0.46 &     2.66 &  &  1.00 &    0.79 &  0.21 &
    0.43 &  &  0.60 &    0.57 &  0.03 &     0.16 &  \\[3pt]
\texttt{SXS:BBH:0214}\\
EOBNR & 32.25 &   30.66 &  1.59 &     4.99 &  &  1.00 &    0.78 &  0.22 &
    0.40 &  & -0.44 &   -0.52 &  0.08 &     0.30 &  \\
IMRPhenom  & 32.25 &   31.64 &  0.61 &     4.92 &  &  1.00 &    0.80 &  0.20 &
    0.41 &  & -0.44 &   -0.46 &  0.02 &     0.29 &  \\[3pt]
\texttt{SXS:BBH:0311}\\
EOBNR & 32.11 &   32.32 & -0.21 &     3.47 &  &  0.84 &    0.79 &  0.05 &
    0.44 &  &  0.40 &    0.38 &  0.02 &     0.23 &  \\
IMRPhenom  & 32.11 &   31.91 &  0.20 &     3.14 &  &  0.84 &    0.78 &  0.06 &
    0.44 &  &  0.40 &    0.39 &  0.02 &     0.19 &  \\[3pt]
\texttt{SXS:BBH:0310}\\
EOBNR & 32.06 &   31.43 &  0.63 &     5.54 &  &  0.82 &    0.71 &  0.11 &
    0.44 &  &  0.00 &   -0.04 &  0.04 &     0.34 &  \\
IMRPhenom  & 32.06 &   31.85 &  0.21 &     4.46 &  &  0.82 &    0.80 &  0.02 &
    0.40 &  &  0.00 &   -0.02 &  0.02 &     0.25 &  \\[3pt]
\texttt{SXS:BBH:0309}\\
EOBNR & 32.06 &   31.22 &  0.85 &     5.27 &  &  0.82 &    0.73 &  0.09 &
    0.43 &  & -0.02 &   -0.05 &  0.04 &     0.34 &  \\
IMRPhenom  & 32.06 &   31.70 &  0.36 &     4.11 &  &  0.82 &    0.81 &  0.00 &
    0.38 &  & -0.02 &   -0.02 &  0.01 &     0.24 &  \\[3pt]
\texttt{SXS:BBH:0305}\\
EOBNR & 32.06 &   31.27 &  0.79 &     5.48 &  &  0.82 &    0.72 &  0.10 &
    0.43 &  & -0.02 &   -0.06 &  0.04 &     0.34 &  \\
IMRPhenom  & 32.06 &   31.79 &  0.27 &     4.16 &  &  0.82 &    0.81 &  0.01 &
    0.40 &  & -0.02 &   -0.02 &  0.01 &     0.24 &  \\[3pt]
\texttt{SXS:BBH:0307}\\
EOBNR & 32.05 &   31.32 &  0.73 &     4.59 &  &  0.81 &    0.73 &  0.09 &
    0.40 &  & -0.08 &   -0.12 &  0.04 &     0.31 &  \\
IMRPhenom  & 32.05 &   31.58 &  0.47 &     4.12 &  &  0.81 &    0.83 & -0.01 &
    0.38 &  & -0.08 &   -0.10 &  0.01 &     0.25 &  \\[3pt]
\texttt{SXS:BBH:0019}\\
EOBNR & 31.47 &   32.24 & -0.77 &     6.19 &  &  0.67 &    0.69 & -0.03 &
    0.45 &  & -0.10 &   -0.10 &  0.00 &     0.37 &  \\
IMRPhenom  & 31.47 &   32.66 & -1.19 &     5.40 &  &  0.67 &    0.75 & -0.08 &
    0.46 &  & -0.10 &   -0.08 & -0.02 &     0.27 &  \\[3pt]
\texttt{SXS:BBH:0239}\\
EOBNR & 30.05 &   31.31 & -1.26 &     7.93 &  &  0.50 &    0.52 & -0.02 &
    0.50 &  &  0.04 &    0.02 &  0.02 &     0.40 &  \\
IMRPhenom  & 30.05 &   29.96 &  0.09 &     6.70 &  &  0.50 &    0.46 &  0.04 &
    0.35 &  &  0.04 &   -0.04 &  0.07 &     0.30 &  \\[3pt]
\texttt{SXS:BBH:0257}\\
EOBNR & 30.05 &   28.84 &  1.21 &     3.97 &  &  0.50 &    0.35 &  0.15 &
    0.69 &  &  0.85 &    0.86 & -0.01 &     0.18 &  \\
IMRPhenom  & 30.05 &   29.96 &  0.09 &     2.17 &  &  0.50 &    0.51 & -0.01 &
    0.27 &  &  0.85 &    0.84 &  0.01 &     0.15 &  \\[3pt]
\texttt{SXS:BBH:0233}\\
EOBNR & 30.05 &   32.80 & -2.75 &    12.00 &  &  0.50 &    0.64 & -0.14 &
    0.60 &  & -0.30 &   -0.32 &  0.02 &     0.57 &  \\
IMRPhenom  & 30.05 &   29.56 &  0.49 &     9.37 &  &  0.50 &    0.44 &  0.06 &
    0.40 &  & -0.30 &   -0.39 &  0.09 &     0.38 &  \\[3pt]
\texttt{SXS:BBH:0065}\\
EOBNR & 18.47 &   18.08 &  0.39 &     1.17 &  &  0.12 &    0.13 & -0.00 &
    0.05 &  &  0.44 &    0.39 &  0.05 &     0.11 &  \\
IMRPhenom  & 18.47 &   18.51 & -0.04 &     1.24 &  &  0.12 &    0.13 & -0.00 &
    0.04 &  &  0.44 &    0.46 & -0.01 &     0.10 &  \\
\end{tabular}

  \caption{True values, medians, absolute biases (difference between true value
and the median) and the width of $90\%$ credible intervals for several
additional aligned-spin \NR
configurations (see Table~\ref{tbl:NRparams} for the parameters of \texttt{SXS:BBH:0307} and Table~\ref{tbl:NRparams_additional} for all other \NR simulation parameters). The results are given for the
chirp mass $\mathcal{M}$, the mass-ratio $q$, and the effective aligned spin
$\chi_\mathrm{eff}$. The \NR waveforms are injected at fiducial inclination angle $\iota=163\degree$, and parameter estimation is performed using the non-precessing
EOBNR and the non-precessing IMRPhenom models.
  }
  \label{tab:aligned_summary}
  \end{ruledtabular}
\end{table*}

\begin{table*}[t]
  \centering
\begin{ruledtabular}
  \begin{tabular}{@{}lrrrrrrrrrrrrrrrr@{}}
  & \multicolumn{3}{c}{$\mathcal{M} ~~ (M_\odot)$} & \phantom{abc}&
\multicolumn{3}{c}{$q $} & \phantom{abc}&
\multicolumn{3}{c}{$\chi_\mathrm{eff}$} & \phantom{abc}&
\multicolumn{3}{c}{$\chi_\mathrm{p}$}\\
  {} &  Median &  Bias &  90\% CI &        &  Median &  Bias &  90\% CI &
&  Median &  Bias &  90\% CI &        &  Median &  Bias &  90\% CI &
\\\hline
\texttt{SXS:BBH:0049} & \multicolumn{3}{c}{$\mathcal{M} = 27.15$~$M_\odot$} & &
\multicolumn{3}{c}{$q = 0.3$} &&  \multicolumn{3}{c}{$\chi_\mathrm{eff} =
0.13$} && \multicolumn{3}{c}{$\chi_\mathrm{p} = 0.5$}\\
  $\iota = 163\degree$                      &   27.47 & -0.32 &     4.92 &  &    0.31 &  0.02 &     0.18 &  &    0.14 & -0.01 &     0.24 &  &    0.20 &  0.30 &     0.46 &  \\
  $\iota = 90\degree$                     &   20.28 &  6.87 &     3.44 &  &    0.28 &  0.05 &     0.12 &  &   -0.66 &  0.78 &     0.28 &  &    0.14 &  0.36 &     0.13 &  \\
  $\iota = 90\degree, \, \psi=120\degree$ &   29.06 & -1.92 &     6.28 &  &    0.33
&  0.01 &     0.14 &  &    0.19 & -0.06 &     0.33 &  &    0.60 & -0.10 &
0.28 &  \\[8pt]
\texttt{SXS:BBH:0522} & \multicolumn{3}{c}{$\mathcal{M} = 30.79$~$M_\odot$} & &
\multicolumn{3}{c}{$q = 0.57$} &&  \multicolumn{3}{c}{$\chi_\mathrm{eff} =
-0.65$} && \multicolumn{3}{c}{$\chi_\mathrm{p} = 0.56$}\\
  $\iota = 163\degree$                      &   32.63 & -1.84 &     5.21 &  &    0.79 & -0.22 &     0.42 &  &   -0.56 & -0.09 &     0.30 &  &    0.39 &  0.17 &     0.50 &  \\
  $\iota = 90\degree$                     &   30.26 &  0.53 &     9.46 &  &    0.46 &  0.11 &     0.58 &  &   -0.55 & -0.11 &     0.46 &  &    0.36 &  0.20 &     0.59 &  \\
  $\iota = 90\degree, \, \psi=120\degree$ &   31.06 & -0.27 &     5.98 &  &    0.67
& -0.10 &     0.49 &  &   -0.63 & -0.03 &     0.35 &  &    0.39 &  0.17 &
0.48 &  \\[8pt]
\texttt{SXS:BBH:0531} & \multicolumn{3}{c}{$\mathcal{M} = 30.8$~$M_\odot$} & &
\multicolumn{3}{c}{$q = 0.57$} &&  \multicolumn{3}{c}{$\chi_\mathrm{eff} =
0.66$} && \multicolumn{3}{c}{$\chi_\mathrm{p} = 0.55$}\\
  $\iota = 163\degree$                      &   30.29 &  0.51 &     3.08 &  &    0.46 &  0.11 &     0.35 &  &    0.61 &  0.05 &     0.19 &  &    0.36 &  0.19 &     0.45 &  \\
  $\iota = 90\degree$                     &   27.06 &  3.73 &     4.26 &  &    0.25 &  0.32 &     0.13 &  &    0.50 &  0.16 &     0.22 &  &    0.38 &  0.17 &     0.37 &  \\
  $\iota = 90\degree, \, \psi=120\degree$ &   30.51 &  0.29 &     3.32 &  &    0.44 &  0.13 &     0.34 &  &    0.63 &  0.03 &     0.20 &  &    0.29 &  0.26 &     0.44 &  \\
  \end{tabular}
\end{ruledtabular}
  \caption{
  Medians, absolute biases (difference between injected value and the median) and the width of $90\%$ credible intervals for several SXS configurations (see Table~\ref{tbl:NRparams_additional}). The results are given for the chirp mass
$\mathcal{M}$, the mass-ratio $q$, the effective aligned spin $\chi_\mathrm{eff}$ and
the effective precession spin $\chi_p$. The precessing IMRPhenom model was used as a template for the fiducial
inclination and edge-on inclination $\iota = 90\degree$. The polarization angle is fixed to the fiducial value $\psi \sim 82\degree$, except where indicated.
  }
  \label{tab:precessing_summary}
\end{table*}

\section{Distribution of detectable polarization and inclination}
\label{sec:angle_distribution}

Here we provide a simple, but instructive estimate of how many observations
are expected to fall into a given range of orientations (i.e., have
particular polarization and inclination angles). This question arose in
Sec.~\ref{sec:effects_of_pol} where the signals with largest bias were found
to be characterized by a specific orientation.
Our analysis follows because the binary, with so little time to precess in band,
can be reasonably approximated by a non-precessing binary for the time interval
of greatest interest.

We start from the detector response, cf.\ Eq.~(\ref{eq:detector_response}),
and the antenna response functions $F_{+,\times}$ that
depend on the polarization-angle $\psi$ and sky-location.
In an Earth-centered coordinate system, the position of the \GW source
on the celestial sphere is given by the spherical polar coordinates $(\beta, \phi)$,
where $\beta$ is related to the declination $\delta$ and $\phi$ to the right
ascension $\alpha$ (see~\cite{anderson:2000yy} for details). The antenna response
functions then read as
\begin{align}
\begin{split}
  F_+ &= -\frac{1+ \cos^2 \beta}{2} \cos 2\phi \, \cos 2\psi -
\cos \beta \sin 2\phi \, \sin 2\psi,  \\
F_\times &= \frac{1+ \cos^2 \beta}{2} \cos 2\phi \, \sin 2\psi
- \cos \beta \sin 2\phi \,  \cos 2\psi.
\end{split} \label{eq:resp_func_standard}
\end{align}
Trigonometric identities allow us to recast (\ref{eq:resp_func_standard})
into the following form
\begin{align}
\begin{split}
F_+ &= - A_{\rm sky} \cos \left(2 \psi - \Xi \right), \\
 F_\times &= A_{\rm sky} \sin \left(2 \psi - \Xi \right),
\end{split} \label{eq:resp_func_geom}
\end{align}
where
\begin{align}
 A_{\rm sky} &  = \sqrt{ \frac{(1+\cos^2 \beta)^2}{4}
\cos^2 2\phi + \cos^2 \beta \, \sin^2 2\phi}, \\
  \Xi   &=  \arctan \left( \frac{2 \cos
\beta}{1+\cos^2 \beta} \tan 2\phi \right).
\end{align}

For non-precessing
binaries and to lowest \PN order, the source inclination, $\iota$, enters the
amplitude of the \GW polarizations $h_{+,\times}$ in the following way~\cite{sathyaschutzlrr},
\begin{align}
\begin{split}
 h_+ &= (1 + \cos^2 \iota) \; A_{\rm GW}  \cos(\phi_{\rm GW}), \\
 h_\times &= -2 \cos \iota \; A_{\rm GW}  \sin(\phi_{\rm GW}).
 \end{split} \label{eq:h_incl}
\end{align}
Using (\ref{eq:resp_func_geom}) and (\ref{eq:h_incl}), we can now recast the
detector response Eq. (\ref{eq:detector_response}) as
\begin{equation}
h_\mathrm{resp} = A_{\rm GW} \, A_{\rm sky} \, A_{\rm pol} \, \cos(\Phi_{\rm GW} -
\Phi_0). \label{eq:det_resp_geom}
\end{equation}
Here, $A_{\rm GW}$ depends on
the
binary's masses, spins and time; $A_{\rm sky}$ depends solely on
the sky
location, and $A_{\rm pol}$ describes the amplitude variation with
inclination and polarization,
\begin{equation}
 A_{\rm pol}  = \sqrt{\left( 1+
\cos^2 \iota \right)^2  \cos^2 (2\psi - \Xi)  + 4 \cos^2 \iota \, \sin^2 (
2\psi - \Xi) }.
\end{equation}
$\Phi_0$ is a simple shift in the phase of $h$,
\begin{equation}
 \Phi_0 = \pi + \arctan \left[\frac{2 \cos \iota}{1 + \cos^2 \iota} \tan(2\psi
- \Xi) \right].
\end{equation}

We now assume that signals with an \SNR above an arbitrary threshold
are detectable. The \SNR is proportional to the signal amplitude which in turn
scales linearly with the inverse of the distance between source and detector.
Assuming uniformly distributed sources, the number of detectable signals is
proportional to the cubed distance, hence we can integrate $A^3_{\rm
pol}$ [all other amplitude terms in (\ref{eq:resp_func_geom}) are constant]
over
polarization, $\psi$, and inclination, $\cos \iota$ (using isotropic priors),
to estimate how
many observations would fall into a particular range of source
orientations.

We find that there is only a 0.3\% chance of a detectable signal to fall
into a $30^\circ \times 30^\circ$ region in inclination and polarization
around the point of minimal amplitude (which we take approximately as
the point of maximal bias).

We stress that this estimate relies on leading order expansions of the
amplitude and assumes a fixed region in inclination-polarization space,
independently of the \SNR. We can drop the first assumption by repeating the
calculation with precessing \NR waveforms, and we find comparable results.
However, whether or not sources show biased
parameter estimates (the original question posed in
Sec.~\ref{sec:effects_of_pol}) depends of course not only on the orientation,
but on the intrinsic parameters and the \SNR of the source; exploring these
parameter dependencies is a long-term goal requiring many more simulation and
analysis campaigns. What we have presented here is an illustration with basic
calculations that only a small fraction of observable sources is expected to
be in the most problematic region of orientations.

\bibliography{NRARPE}

\clearpage
\begin{widetext}

\centerline{\large \textbf{Authors}}

\overfullrule 0pt
\parskip0pt
\hyphenpenalty9999

\noindent
B.~P.~Abbott,$^{1}$  
R.~Abbott,$^{1}$  
T.~D.~Abbott,$^{2}$  
M.~R.~Abernathy,$^{3}$  
F.~Acernese,$^{4,5}$ 
K.~Ackley,$^{6}$  
C.~Adams,$^{7}$  
T.~Adams,$^{8}$ 
P.~Addesso,$^{9}$  
R.~X.~Adhikari,$^{1}$
% ...
\\

\medskip
\parindent 0pt
{${}^{*}$Deceased, March 2016. }%
{${}^{**}$Deceased, March 2017. }%
\medskip

$^{1}$LIGO, California Institute of Technology, Pasadena, CA 91125, USA 

$^{2}$Louisiana State University, Baton Rouge, LA 70803, USA 

$^{3}$American University, Washington, D.C. 20016, USA 

$^{4}$Universit\`a di Salerno, Fisciano, I-84084 Salerno, Italy 

$^{5}$INFN, Sezione di Napoli, Complesso Universitario di Monte S.Angelo, I-80126 Napoli, Italy 

$^{6}$University of Florida, Gainesville, FL 32611, USA 

$^{7}$LIGO Livingston Observatory, Livingston, LA 70754, USA 

$^{8}$Laboratoire d'Annecy-le-Vieux de Physique des Particules (LAPP), Universit\'e Savoie Mont Blanc, CNRS/IN2P3, F-74941 Annecy-le-Vieux, France 

$^{9}$University of Sannio at Benevento, I-82100 Benevento, Italy and INFN, Sezione di Napoli, I-80100 Napoli, Italy 

% ...

\end{widetext}

\end{document}